\documentclass[11pt]{article}
\pdfoutput=1

\usepackage{cancel,slashed}
\usepackage{bbm}
\usepackage{amssymb}
\usepackage{amsfonts}
\usepackage{amsmath}
\usepackage{graphicx}
\usepackage{cite}
\usepackage{physics}

\usepackage{ dsfont }
\usepackage{latexsym}
\usepackage{color}
\usepackage{cancel,slashed}
\usepackage[hyperfootnotes=false,linktocpage]{hyperref}
\usepackage{color}
\usepackage{transparent}
\usepackage[hang,flushmargin]{footmisc}

\usepackage{blkarray}
\usepackage{multirow}
\usepackage{empheq}
\usepackage{caption}
\usepackage{subcaption}
\usepackage{comment}

\newcommand{\bmat}{\left(\begin{array}}
\newcommand{\emat}{\end{array}\right)}

\def\Z{\mathbb{Z}}

\def\P{\mathbb{P}}
\def\CK {{\cal K}}
\def\a {\alpha}
\def\b {\beta}

\def\1{{\bf 1}}
\def\2{{\bf 2}}
\def\3{{\bf 3}}
\def\4{{\bf 4}}
\def\6{{\bf 6}}

\def\targ#1#2{\genfrac{[}{]}{0pt}{}{#1}{#2}}
\def\targ2#1#2{\genfrac{}{}{0pt}{}{#1}{#2}}

\definecolor{mygr}{rgb}{0,0.6,0}
\definecolor{mygrey}{rgb}{0,0.1,0.2}
\definecolor{myblue}{rgb}{0,0.5,0.9}
\definecolor{myblue2}{rgb}{0,0.5,0.5}
\definecolor{myblue3}{rgb}{0,0.7,0.9}
\definecolor{myblue4}{rgb}{0,0.6,0.6}
\definecolor{myorange}{rgb}{1,0.5,0}
\definecolor{mypurple}{rgb}{0.6,0,1}
\definecolor{mygolden}{rgb}{1,0.8,0.2}
\definecolor{mycyan}{rgb}{0,1,1}
\definecolor{mymagenta}{rgb}{1,0,1}
\definecolor{mykiwi}{rgb}{0.8,1,0.5}
\definecolor{mybrown}{cmyk}{0.14, 0.42, 0.56, 0.2}
\definecolor{myturq}{cmyk}{0.99, 0, 0.2, 0.4}
\definecolor{myaubergine2}{cmyk}{0.4, 0.5, 0, 0.1}
\definecolor{myaubergine}{cmyk}{0.6,0.85,0,0}
\definecolor{CycleGreen}{cmyk}{0.52,0,1,0}
\definecolor{CycleBrown}{cmyk}{0, 0.4, 0.9, 0.2}

\DeclareFontFamily{U}{rcjhbltx}{}
\DeclareFontShape{U}{rcjhbltx}{m}{n}{<->rcjhbltx}{}
\DeclareSymbolFont{hebrewletters}{U}{rcjhbltx}{m}{n}

\DeclareMathSymbol{\lamed}{\mathord}{hebrewletters}{108}
\DeclareMathSymbol{\mem}{\mathord}{hebrewletters}{109}
\DeclareMathSymbol{\ayin}{\mathord}{hebrewletters}{96}
\DeclareMathSymbol{\tsadi}{\mathord}{hebrewletters}{118}
\DeclareMathSymbol{\qof}{\mathord}{hebrewletters}{113}
\DeclareMathSymbol{\resh}{\mathord}{hebrewletters}{114}
\DeclareMathSymbol{\pe}{\mathord}{hebrewletters}{112}
\DeclareMathSymbol{\pesofit}{\mathord}{hebrewletters}{80}
\DeclareMathSymbol{\samekh}{\mathord}{hebrewletters}{115}
\DeclareMathSymbol{\tav}{\mathord}{hebrewletters}{116}
\DeclareMathSymbol{\vav}{\mathord}{hebrewletters}{119}
\DeclareMathSymbol{\het}{\mathord}{hebrewletters}{120}
\DeclareMathSymbol{\yod}{\mathord}{hebrewletters}{121}
\DeclareMathSymbol{\zayin}{\mathord}{hebrewletters}{122}
\DeclareMathSymbol{\alephdot}{\mathord}{hebrewletters}{128}
\DeclareMathSymbol{\tsadisofit}{\mathord}{hebrewletters}{90}
\DeclareMathSymbol{\shin}{\mathord}{hebrewletters}{152}

\def\CN {{\cal N}}

\def\d{{\delta}}
\def\be{\begin{equation}}
\def\ee{\end{equation}}
\def\bea{\begin{eqnarray}}
\def\eea{\end{eqnarray}}
\def\bes{\begin{subequations}}
\def\ees{\end{subequations}}

\def\eps{{\epsilon}}
\def\oh{\frac{1}{2}}

\def\re{\mbox{Re}\, }
\def\im{\mbox{Im}\, }

\def\cy {{\text{CY}}}
\def\IZ{\mathbb{Z}}

\def\Om{\Omega}
\usepackage{multicol}
\usepackage{float}
\usepackage{caption}
\def\p {{\partial}}

\def\g {{\gamma}}

\newcommand{\cF}{\mathcal{F}}

\newcommand{\cK}{\mathcal{K}}

\newcommand{\cN}{\mathcal{N}}
\newcommand{\cO}{\mathcal{O}}

\newenvironment{eqn*}{\begin{equation*}\begin{aligned}}{\end{aligned}\end{equation*}\noindent}

\makeatletter
\newsavebox\myboxA
\newsavebox\myboxB
\newlength\mylenA

\newcommand*\xoverline[2][0.75]{%
\sbox{\myboxA}{$\m@th#2$}%
\setbox\myboxB\null
\ht\myboxB=\ht\myboxA%
\dp\myboxB=\dp\myboxA%
\wd\myboxB=#1\wd\myboxA
\sbox\myboxB{$\m@th\overline{\copy\myboxB}$}
\setlength\mylenA{\the\wd\myboxA}
\addtolength\mylenA{-\the\wd\myboxB}%
\ifdim\wd\myboxB<\wd\myboxA%
   \rlap{\hskip 0.5\mylenA\usebox\myboxB}{\usebox\myboxA}%
\else
    \hskip -0.5\mylenA\rlap{\usebox\myboxA}{\hskip 0.5\mylenA\usebox\myboxB}%
\fi}
\makeatother

\begin{document}
\pagestyle{plain}

\makeatletter
\@addtoreset{equation}{section}
\makeatother
\renewcommand{\theequation}{\thesection.\arabic{equation}}

\pagestyle{empty}
\rightline{IFT-UAM/CSIC-22-46}
\vspace{0.5cm}
\begin{center}
\Huge{{Membranes in AdS$_4$ orientifold vacua \\  and their Weak Gravity Conjecture}
\\[10mm]}
\normalsize{Gonzalo F. Casas, Fernando Marchesano, and  David Prieto \\[12mm]}
\small{
Instituto de F\'{\i}sica Te\'orica UAM-CSIC, c/ Nicol\'as Cabrera 13-15, 28049 Madrid, Spain
\\[10mm]} 
\small{\bf Abstract} \\[5mm]
\end{center}
\begin{center}
\begin{minipage}[h]{15.0cm} 

We study type IIA orientifold compactifications with fluxes that give rise to perturbatively stable, non-supersymmetric AdS$_4$ vacua with D6-brane gauge sectors. Non-perturbative instabilities can be mediated by D8-branes wrapped on the six internal dimensions $X_6$, if they reduce to 4d membranes with a charge $Q$ larger than its tension $T$. The mismatch $Q \neq T$ arises due to {\it i)} curvature corrections and {\it ii)} the BIon backreaction of D6-branes wrapping 3-cycles of $X_6$. We give a simple expression for the second effect in toroidal orientifolds, and find that only pairs of 3-cycles at SU(2) angles contribute to it.  They either contribute towards  $Q>T$ or $Q<T$ depending on the 3-cycles separation, allowing to engineer 4d $\cN=0$ vacua in tension with the Weak Gravity Conjecture for membranes.

\end{minipage}
\end{center}
\newpage
\setcounter{page}{1}
\pagestyle{plain}
\renewcommand{\thefootnote}{\arabic{footnote}}
\setcounter{footnote}{0}


\tableofcontents

\section{Introduction and summary}
\label{s:intro}

In order to properly describe the string Landscape one not only needs to provide the set of string vacua, but also specify some key properties like their stability. In this sense, the AdS Instability Conjecture \cite{Ooguri:2016pdq,Freivogel:2016qwc}, that proposes that all $\cN=0$ AdS$_d$ vacua are at best metastable, is a very powerful statement. The proposal is partially motivated by a refinement of the Weak Gravity Conjecture (WGC) applied to $(d-2)$-branes, stating that the WGC inequality is only saturated in supersymmetric setups \cite{Ooguri:2016pdq}. In non-supersymmetric ones, and in particular in AdS$_d$ vacua supported by $d$-form background fluxes, there should be a superextremal $(d-2)$-brane that nucleates and expands towards the AdS$_d$ boundary \cite{Maldacena:1998uz}, mediating a non-perturbative decay. 

This sort of nucleation has been observed in many different contexts, including type II string flux compactifications to AdS \cite{Gaiotto:2009mv,Antonelli:2019nar,Apruzzi:2019ecr,Bena:2020xxb,Suh:2020rma,Guarino:2020jwv,Guarino:2020flh,Basile:2021vxh,Apruzzi:2021nle,Bomans:2021ara}, but it is particularly meaningful in those setups where the AdS scale is much lower than the compactification scale, as it allows us to connect with our standard picture of the string Landscape. This  highlights the massive type IIA orientifold compactifications to AdS$_4$ put forward in \cite{DeWolfe:2005uu,Camara:2005dc}, known in the literature as DGKT-like vacua. These models feature O6-planes and D6-branes wrapping three-cycles of the internal manifold $X_6$, and a set of fluxes generating a 4d potential that is particularly simple in the large volume regime \cite{Bielleman:2015ina,Carta:2016ynn,Herraez:2018vae}. This simplicity allows one to perform a rather general analysis of the set of vacua even when $X_6$ corresponds to a general Calabi--Yau manifold \cite{Escobar:2018tiu,Escobar:2018rna,Marchesano:2019hfb}. 
 From this analysis one encounters a family of supersymmetric vacua and two universal families of non-supersymmetric vacua. Finally, by computing the 4d flux induced spectrum and using scale separation one concludes that the latter are perturbatively stable \cite{Marchesano:2019hfb}. It thus remains to check their non-perturbative stability. 

Given that these are examples of AdS$_4$ vacua supported by 4d fluxes, the proposal of \cite{Ooguri:2016pdq} gives clear candidates to mediate non-perturbative decays, namely 4d membranes coupled to such fluxes, with a charge $Q$ and tension $T$ such that $Q >T$. The most obvious case are D4-branes wrapping (anti)holomorphic two-cycles of $X_6$, in the family of non-supersymmetric vacua that are related to the supersymmetric ones by a sign flip of the internal four-form flux. The membrane charge and tension for such D4-branes were computed in the probe approximation in \cite{Aharony:2008wz,Narayan:2010em}, for both supersymmetric vacua and the said  non-supersymmetric cousins. It was found that in both cases the 4d membranes satisfy the BPS relation $Q=T$, that corresponds to a marginal decay and not to an actual instability. However, such a computation uses a Calabi--Yau metric for $X_6$, which is only an approximation of the actual background. A more precise background that takes into account the backreaction of localised sources can be found following \cite{Junghans:2020acz,Marchesano:2020qvg}. Performing the computation in this new background can be understood as including one-loop corrections to the charge and tension of the membrane. This exercise was made in \cite{Marchesano:2021ycx}, finding that $Q=T$ also at this level. Therefore, checking the proposal of \cite{Ooguri:2016pdq} in this case would require a more accurate description of the background or perhaps alternative techniques as in \cite{Giri:2021eob}, and as of today it remains an open problem. 

Nevertheless, it was argued in \cite{Marchesano:2021ycx} that a second kind of 4d membranes exist which mediates non-perturbative decays. These new membranes are made of D8-branes wrapping the internal manifold $X_6$, and have space-time filling D6-branes attached to them. At leading order they satisfy the BPS equality $Q=T$, but at the level of one-loop corrections and for $\cN =0$ vacua this is no longer true, there being two sources of correction to this equality. The first source is the correction to the D8-brane worldvolume action due to the curvature of $X_6$, that induces a negative D4-brane charge and tension specified by the second Chern class of $X_6$. For the $\cN =0$ vacua of interest this correction is such that $\Delta^{\rm curv}_{\rm D8} (Q-T) > 0$, favouring the nucleation of the membrane towards the AdS$_4$ boundary. The second correction is harder to compute, as it involves the worldvolume flux induced by the BIon-like  backreaction of localised objects, namely the D6-branes ending on the D8-brane. In spite of this, such a correction was computed for the geometry $X_6 = T^6/(\IZ_2 \times \IZ_2)$ assuming a naive, symmetric distribution of D6-branes, finding that  $\Delta^{\rm BIon}_{\rm D8} (Q-T) > 0$. As a result, in this particular case one finds that $Q>T$ and the conjecture of \cite{Ooguri:2016pdq} is verified for D8-branes. If it happened that $\Delta^{\rm BIon}_{\rm D8} (Q-T) > 0$ for arbitrary geometries and D6-brane configurations, one could extend this statement to all $\cN=0$ vacua of this sort, or in other words verify the refined WGC for 4d membranes made out of D8-branes. 

In this work we undertake a more general study of  $\Delta^{\rm BIon}_{\rm D8}$, considering orientifolds of the form $X_6 = (T^2 \times T^2 \times T^2)/\Gamma$ with different orbifold groups $\Gamma$ and D6-brane configurations. Remarkably, we find that for certain D6-brane configurations $\Delta^{\rm BIon}_{\rm D8}  < 0$, even in the simple geometry $X_6 = T^6/(\IZ_2 \times \IZ_2)$. The key ingredient to achieve this negative sign seems to be the presence of localised sources that do not intersect, and in particular non-intersecting O6-planes. 

Indeed, from our analysis one can derive some lessons that we expect to be valid in more general geometries. First, $\Delta^{\rm BIon}_{\rm D8}$ can be split into several contributions, one per each pair of localised sources, or in other words by a pair of three-cycles $\Pi_\a$ and $\Pi_\b$ wrapped by D6-branes, that can always be placed on top of O6-planes. Second, if these two three-cycles have a non-vanishing intersection number or lie in homology classes that are proportional to each other, then the pair does not contribute to $\Delta^{\rm BIon}_{\rm D8}$. Non-vanishing contributions occur instead when $\Pi_\a$ and $\Pi_\b$ are related by a $SU(2)$ rotation in $X_6$. This includes cases where they intersect over a one-cycle, and cases where they do not intersect at all. In these cases the open-string spectrum between $\Pi_\a$ and $\Pi_\b$ typically arranges itself in $\cN=2$ multiplets, and so we dub these kind of pairs as $\cN=2$ sectors of the compactification. Third, in the simple case where D6-branes are placed on top of O6-planes we obtain the simple formula
\begin{equation}
     \Delta_{\rm D8}^{\rm Bion} = \frac{1}{24N_\Gamma} \sum_{(\a,\b)\in \cN=2} \hat{q}_\a \hat{q}_\b\varepsilon_{\a\b} \, \#(\Pi_\a \cap \Pi_\b)_a T^a_{\rm D4}\, .
     \label{introDelta}
\end{equation}
Here $\a$, $\b$ run over the different smooth three-cycles that appear in the covering space $T^6$, which in our setup are nothing but the O6-plane locations $\Pi_{\rm O6} = \cup_\a \Pi_\a$, and $\hat{q}_\a \in \mathbb{N}$ is  the jump in the number of D6-branes wrapped on $\Pi_\a$ when we cross a single  D8-brane. The sum only selects those pairs that are related by an $SU(2)$ rotation: $\#(\Pi_\a \cap \Pi_\b)_a \geq 0$ is the `number of intersections' on the $(T^2)_b \times (T^2)_c$ where this rotation happens and $T^a_{\rm D4}$ is the tension of a D4-brane wrapping the remaining $(T^2)_a$. In particular, if $\Pi_\a$ and $\Pi_\b$ intersect over one-cycles over $(T^2)_a$, then $\#(\Pi_\a \cap \Pi_\b)_a$ is the number of such intersections. Finally, $N_\Gamma$ is the order of the orbifold group and $\varepsilon_{\a\b}$ is an integer that distinguishes between the two possibilities that we encounter in our examples. If $\Pi_\a$ and $\Pi_\b$ intersect over one-cycles then $\varepsilon_{\a\b} =2$, and if they do not intersect at all then $\varepsilon_{\a\b} =-1$. In other words, D6-branes on top of O6-planes that intersect contribute towards $\Delta^{\rm BIon}_{\rm D8} > 0$, and therefore the non-perturbative instability of the vacuum, while those on top of O6-planes that do not intersect have the opposite contribution. 

It follows that if one considers D6-branes configurations with $\cN=2$ pairs where none of them intersect, as it is possible in blown-up orbifold geometries, one necessarily has that $\Delta^{\rm BIon}_{\rm D8} < 0$. Presumably, the same could happen in more general Calabi--Yau geometries where O6-planes do not intersect. In those cases, one needs to insure that $\Delta^{\rm curv}_{\rm D8} > |\Delta^{\rm BIon}_{\rm D8}|$ in order to satisfy the refined WGC for 4d membranes made out of D8-branes. While in most of the examples that we have analysed this is the case, one can also engineer vacua in which $\Delta^{\rm curv}_{\rm D8} + \Delta^{\rm BIon}_{\rm D8} < 0$. Notice that this does  not determine the non-perturbative stability of the $\cN=0$ vacuum, as nucleation of D4-branes wrapping internal two-cycles could still be favoured, but it nevertheless selects  potential counterexamples to the WGC for 4d membranes, up to some caveats that we comment on. It would be important to establish whether the WGC is violated or not for a subfamily of type IIA flux vacua, as this could affect our picture of the string Landscape. 

The rest of the paper is organised as follows. In section \ref{s:nonsusy} we review the AdS$_4$ compactifications of interest and the computation of 4d membrane charges and tensions in them. In section \ref{s:Delta} we summarise how to compute the BIonic excess charge $\Delta^{\rm BIon}_{\rm D8}$ in toroidal orientifolds, based on the explicit computations of section \ref{s:examples}. Given this expression for $\Delta^{\rm BIon}_{\rm D8}$ we provide a simple example in which $\Delta^{\rm curv}_{\rm D8} +\Delta^{\rm BIon}_{\rm D8} < 0$. Due to flux quantisation conditions, such an example must be engineered in a blown-up $T^6/(\IZ_2 \times \IZ_2)$ geometry, discussed in appendix \ref{ap:Z2xZ2}, and whose second Chern class is computed in Appendix \ref{ap:Z2xZ2curv}. We finally draw our conclusions in section \ref{s:conclu}. 


\section{AdS$_4$ orientifold vacua}
\label{s:nonsusy}

To construct non-supersymmetric AdS$_4$ backgrounds let us consider type IIA string theory compactified on a Calabi--Yau three-fold $X_6$. To this background we apply an orientifold quotient generated by $\Omega_p (-1)^{F_L}{\cal R}$, with $\Omega_p$ the worldsheet parity reversal operator, ${F_L}$ the left-movers space-time fermion number and ${\cal R}$ an anti-holomorphic involution of $X_6$, that acts as ${\cal R} J_{\rm CY}=-J_{\rm CY}$ and ${\cal R}\Omega_{\rm CY} = - \overline{\Omega}_{\rm CY}$ on its K\"ahler 2-form and holomorphic 3-form, respectively. The fixed locus $\Pi_{\rm O6}$ of ${\cal R}$ is made of one or several smooth 3-cycles of $X_6$, hosting O6-planes. The presence of O6-planes reduces the background supersymmetry to 4d $\CN=1$, and induces an RR tadpole that can be cancelled by a combination of D6-branes wrapping special Lagrangian three-cycles \cite{Blumenhagen:2005mu,Blumenhagen:2006ci,Marchesano:2007de,Ibanez:2012zz}, D8-branes wrapping coisotropic cycles with fluxes \cite{Font:2006na}, and background fluxes including the Romans mass. For simplicity, in the following we will consider that the D-brane content consists of D6-branes placed on top of the O6-planes or in another representative of the same homology class. The remaining RR tadpole is then cancelled by the presence of backgrounds fluxes, yielding either a 4d $\CN=1$ or $\CN=0$ vacuum. 

In general, the effect of the backgrounds fluxes is two-fold. On the one hand they generate a potential that stabilises the moduli of the Calabi--Yau orientifold compactification, yielding families of supersymmetric and non-supersymmetric vacua. On the other hand they generate a warp factor, a varying dilaton and deform the background away from the Calabi--Yau metric. The first effect was analysed from the four-dimensional viewpoint in \cite{DeWolfe:2005uu,Camara:2005dc} for toroidal geometries, and more recently in \cite{Marchesano:2019hfb} for general Calabi--Yau geometries. The second one was addressed in  \cite{Junghans:2020acz,Marchesano:2020qvg} (see also \cite{Saracco:2012wc,DeLuca:2021mcj}) where the equations of 10d massive type IIA supergravity were expanded in either $g_s$ (the average value of the 10d dilaton $e^{\phi}$) or $\hat{\mu} = \ell_s/R$ (the AdS$_4$ scale in the 10d string frame in units of the string length $\ell_s  =  2\pi \sqrt{\a'}$). The solution for the first terms of this expansion was given quite explicitly in \cite{Marchesano:2020qvg} for supersymmetric vacua, while the same degree of accuracy was extended to one family of non-supersymmetric vacua in \cite{Marchesano:2021ycx}. Finally, it was pointed out in \cite{Marchesano:2021ycx} that 4d membranes  made up from D8-branes wrapped on $X_6$ are natural candidates to mediate a non-perturbative instability in such a family of 4d $\CN=0$ vacua. 

To properly understand this last point let us briefly review some of the results of the above references. One important ingredient is the flux background of these compactifications, that can be conveniently described using the democratic formulation of type IIA supergravity \cite{Bergshoeff:2001pv}, in which all RR potentials are grouped in a polyform ${\bf C} = C_1 + C_3 + C_5 + C_7 + C_9$ and so are their gauge invariant field strengths
\be
{\bf G} \,=\, d_H{\bf C} + e^{B} \wedge {\bf \bar{G}} \, .
\label{bfG}
\ee
Here $H$ is the three-form NS flux, $d_H \equiv (d - H \wedge)$ is the $H$-twisted differential  and ${\bf \bar{G}}$ a formal sum of closed $p$-forms on $X_6$. The Bianchi identities read
\begin{equation}\label{IIABI}
\ell_s^{2} \,  d (e^{-B} \wedge {\bf G} ) = - \sum_\a \lambda \left[\delta (\Pi_\alpha)\right] \wedge e^{\frac{\ell_s^2}{2\pi} F_\alpha} \, ,  \qquad d H = 0 \, ,
\end{equation} 
where $\Pi_\alpha$ hosts a D-brane source with a quantised worldvolume flux $F_\alpha$, and $\delta(\Pi_\alpha)$ is the bump $\delta$-function form with support on $\Pi_\alpha$ and indices transverse to it, such that $\ell_s^{p-9} \d(\Pi_\a)$ lies in the Poincar\'e dual class to $[\Pi_\a]$. O6-planes contribute as D6-branes but with minus four times their charge and $F_\alpha \equiv 0$. Finally, $\lambda$ is the operator that reverses the order of the indices of a $p$-form.

In the presence of D6-branes and O6-planes the Bianchi identities for the RR fluxes read
\be
dG_0 = 0\, , \qquad d G_2 = G_0 H - 4 \d_{\rm O6} +   N_\a \d_{\rm D6}^\a \, ,  \qquad d G_4 = G_2 \wedge H\, , \qquad dG_6 = 0\, ,
\label{BIG}
\ee
where  we have defined $\d_{\rm D6/O6}\equiv \ell_s^{-2}  \d(\Pi_{\rm D6/O6})$. This in particular implies that
\be
{\rm P.D.} \left[4\Pi_{\rm O6}- N_\a \Pi_{\rm D6}^\a\right] = m [\ell_s^{-2} H] \, ,
\label{tadpole}
\ee
where we assume that the NS flux $H$ is closed. Here $m = \ell_s G_0 \in \mathbb{Z}$ is the quantum of Romans mass, and  $N_\a$ the number D6-branes wrapping a three-cycle in the  homology class $[\Pi^{\rm D6}_\a]$. 

The 4d  vacua analysis yields the following conditions on the internal background fluxes:
\be
[ H ]  = \frac{2}{5} G_0 g_s  [\re \Omega_{\rm CY} ] \, , \quad \int_{X_6} G_2 \wedge \tilde{\omega}^a =  0\, ,  \quad \frac{1}{\ell_s^6} \int_{X_6} G_4  \wedge \omega_a  =  \epsilon \frac{3}{10} G_0 {\cal K}_a \, , \quad  G_6  =  0\, , 
\label{intflux}
\ee
where $\omega_a$, $\tilde \omega^a$ are integral harmonic two- and four-forms of $X_6$ such that $\ell_s^{-6} \int_{X_6} \omega_a \wedge \tilde{\omega}^b = \delta_a^b$, and $\CK_a = \int_{X_6} J_{\rm CY} \wedge J_{\rm CY} \wedge \omega_a$. Here  $\epsilon = -1$ describes supersymmetric backgrounds, while $\epsilon = 1$ corresponds to non-supersymmetric vacua. Finally we have that
\be
\mu = \frac{1}{R} = \frac{1}{5} |G_0| g_s  \, ,
\label{Rads}
\ee
relating the AdS$_4$ radius with the average 10d dilaton value, which is in turn fixed by the equations \eqref{intflux}, and is a small parameter for those solutions with large Calabi--Yau volume ${\cal V}_{\rm CY}$.

With these data and using pure spinor techniques an approximate solution to the 10d massive type IIA supersymmetry equations was found in \cite{Marchesano:2020qvg}, as the first terms of an expansion in $g_s$. Similarly, following the more general approach of \cite{Junghans:2020acz}, an approximate solution to the 10d equations of motion corresponding to  the non-supersymmetric vacua with $\epsilon=1$ was found in \cite{Marchesano:2021ycx}. Both of these 10d backgrounds display a warped metric of the form
\begin{equation}\label{eq:warped-product}
	ds^2 = e^{2A}ds^2_{\mathrm{AdS}_4} + ds^2_{X_6}\, ,
\end{equation}
 with $A$ a function on $X_6$. The metric on $X_6$ is not Calabi--Yau, but a deformation to a $SU(3)\times SU(3)$ structure metric. Such a deformation is described in terms of a (2,1) primitive current $k$ and a real function $\varphi$ that satisfies $\int_{X_6} \varphi = 0$. More precisely, one obtains the following metric background and dilaton profile
 \begin{subequations}	
	\label{solutionsu3}
\begin{align}
J & = J_{\rm CY} + \cO(g_s^2) \, , \qquad   \qquad  \Omega  = \Omega_{\rm CY} + g_s k +  \cO(g_s^2)\, , \\
e^{-A}  & = 1 + g_s \varphi + \cO(g_s^2) \, , \qquad e^{\phi}   = g_s \left(1 - 3  g_s \varphi\right) + \cO(g_s^3)\, ,
\end{align}
\end{subequations}   
for  $\epsilon = \pm 1$. The precise profile for $\varphi$ and $k$ is found by solving the Bianchi identity \eqref{BIG} for $G_2$. 

Indeed, let us express the RR two-form flux in terms of a three-form current $K$ as $G_2 = d^\dag_{\rm CY} K$, so that its Bianchi identity reads
\begin{equation}
    \Delta_{\rm CY} K = G_0H +  \delta_{\rm O6+D6}    = \frac{2}{5} m^2 g_s \ell_s^{-2} \re \Omega_{\rm CY}+  \delta_{\rm O6+D6}  +  \cO(g_s^2)\, ,
    \label{eq: K equation}
\end{equation}
where we have defined $\Delta_{\rm CY} = d^\dag_{\rm CY} d + d d^\dag_{\rm CY}$ and $\delta_{\rm O6+D6} = - 4\delta_{\rm O6} + N_\a \delta_{\rm D6}^\a$, and we have used  the leading term in the expansion of $H$, see below. This equation has a solution if \eqref{tadpole} is satisfied, and it is particularly simple at leading order in $g_s$ if the D6-branes wrap special Lagrangian three-cycles $\Pi^{\rm D6}_\a$ that are mutually BPS with $\Pi_{\rm O6}$. At this level $H$ is a harmonic three-form, which means that we can decompose the leading term of the rhs of \eqref{eq: K equation} as
\begin{equation}
  \ell_s^{-2} \sum_{\a,\eta} q_{\a, \eta} \left(  H_\a - \delta(\Pi_{\a,\eta}) \right) \, .
    \label{Krhsum}
\end{equation}
Here $\Pi_{\a, \eta}$ is a three-cycle hosting a localised source, either D6-brane or O6-plane, and $q_{\a, \eta} \in \IZ$ minus its charge in D6-brane units. The index $\eta$ labels different three-cycles that correspond to the same homology class: $[\Pi_{\a,\eta}] = [\Pi_{\a}]$, $\forall \eta$.  Finally, $H_\a$ is the harmonic representative of the Poincar\'e dual class to $\ell_s^3 [\Pi_{\a}]$. Then, using that $\Pi_{\a,\eta}$ are special Lagrangian three-cycles calibrated by $\im \Omega_{\rm CY}$, one can show that the Laplace equations
\begin{equation}
    \ell_s^2  \Delta_{\rm CY} K_{\a,\eta} = H_\a - \delta(\Pi_{\a,\eta})\, .
    \label{eq: Kalpha equation}
 \end{equation}
 have a solution of the form \cite{Hitchin:2010qz,Marchesano:2020qvg}
 \be
 K_{\a,\eta} = \varphi_{\a,\eta} \re \Omega_{\rm CY}  + \re k_{\a,\eta} \, ,
\label{formKalpha}
\ee
and by linearity of the equation \eqref{eq: K equation} one can express $K$ as
\be
 K =  \sum_{\a,\eta} q_{\a, \eta} K_{\a,\eta} = \varphi \re \Omega_{\rm CY}  + \re k \, ,
\label{formK}
\ee
and so the quantities $\varphi$ and $k$ that determine the background \eqref{solutionsu3} are given by $ \varphi = \sum_{\a,\eta} q_{\a, \eta} \varphi_{\a,\eta}$ and $ k =  \sum_{\a,\eta} q_{\a, \eta} k_{\a,\eta}$, respectively. In particular we have that
\be
\Delta_{\rm CY}  \varphi_{\a,\eta} = \left(\frac{{\cal V}_{\Pi_\a}}{{\cal V}_{\rm CY}} - \delta^{(3)}_{\a,\eta}\right)  \ \implies \ \varphi \sim \cO(g_s^{1/3})\, ,
\ee
where $\delta^{(3)}_{\a,\eta} \equiv *_{\rm CY} (\im \Omega_{\rm CY} \wedge \delta(\Pi_{\a,\eta}))$, ${\cal V}_{\rm CY} = -\frac{1}{6}\ell_s^{-6} \int_{X_6} J_{\rm CY}^3$ is the Calabi--Yau volume and ${\cal V}_{\Pi^{\rm O6}_\a} = \ell_s^{-3} \int_{\Pi_\a} \im \Omega_{\rm CY}$. As a result $\varphi \sim - \frac{q_{\a, \eta}}{r}$ in the vicinity of a $\Pi^{\rm O6}_{\a,\eta}$. If the localised charge is negative it describes a small region where the 10d string coupling blows up, the warp factor becomes negative and, as expected, the supergravity approximation cannot be trusted. 

Let us consider a simplified setup in which all localised sources wrap three-cycles determined by the O6-plane locus. We describe the O6-plane locus  as a union of several smooth three-cycles
\begin{equation}
    \Pi_{\rm O6} = \bigcup_{\a,\eta} \Pi_{\a,\eta}\, , \quad \text{with} \quad    [\Pi_{\rm O6}] = \sum_\a p_\a [\Pi_{\a}]\, ,
    \label{splitO6}
\end{equation}
where  the index $\alpha$ runs over different homology classes and $\eta$ over the $p_\a$ different representatives of the same homology class: $[\Pi^{\rm O6}_{\a,\eta}] = [\Pi^{\rm O6}_{\a,\eta'}] \equiv [\Pi^{\rm O6}_{\a}]$. Then we consider D6-branes that wrap three-cycles on the same homology classes, that is we take $[\Pi^{\rm D6}_{\a}] = [\Pi^{\rm O6}_{\a}]$. One may further assume that all D6-branes lie on top of O6-planes, so $\Pi^{\rm D6}_{\a,\eta} = \Pi^{\rm O6}_{\a,\eta}$. An advantage of this further simplification is that on top of the O6-planes one can always have a vanishing worldvolume flux for the D6-brane, which is a necessary condition for a vacuum. If we displace such a D6-brane away from the O6-plane location the presence of the $H$-flux will generically induce a $B$-field in its worldvolume, that will generate a dynamical tadpole.\footnote{In general there will be a discretum of other representatives within $ [\Pi^{\rm O6}_{\a}]$ besides the O6-plane locus where the D6-brane worldvolume flux can vanish, similarly to the open string landscape in \cite{Gomis:2005wc}. Our discussion below can be easily extended to include those D6-brane locations as well.} Then, in an analogous fashion to \cite{Mininno:2020sdb}, the WGC could be violated due to the lack of equilibrium. Our choice avoids such a possibility. 

To sum up, we consider a setup in which the three-cycles $\Pi_{\a, \eta}$ in \eqref{Krhsum} correspond to those in \eqref{splitO6}. As a result 
\begin{equation}
   \ell_s^{2} \delta_{\rm O6+D6} = - \sum_{\a,\eta}  q_{\a,\eta} \delta(\Pi^{\rm O6}_{\a,\eta})\, ,
   \label{O6D6source}
\end{equation}
where $q_{\a,\eta} = 4 - N_{\a, \eta}$ is minus the localised charge on each three-cycle. We also choose P.D.$[\ell_s^{-2}H] = h [\Pi_{\rm O6}]$ and $N_\a \equiv \sum_{\eta} N_{\a,\eta}  = N p_\a$, which leads to the simple tadpole constraint
\be
mh = 4 - N  \, .
\label{tadpole2}
\ee
Here notice that $h$ and $N$ need not be integers, because a consistent configuration only requires that $h[\Pi_{\rm O6}]$ and $N [\Pi_{\rm O6}]$ are integer homology classes. So if $[\Pi_{\rm O6}] = M [\hat{\Pi}_{\rm O6}]$, with $M \in \IZ$ and $[\hat{\Pi}_{\rm O6}] \in H_{3}(X_6, \IZ)$, we only need to require that $hM, NM \in \IZ$, as will happen in the toroidal orientifold geometries that we will analyse in the following sections. Additionally, the 4d analysis on vacua conditions requires that $mh$ and $N, N_{\a,\eta}$ are non-negative, so that there is a finite number of solutions to the tadpole equation. 

The approximate flux background is also described in terms of $\varphi$ and $k$. We have that
\begin{subequations}
	\label{solutionflux}
\begin{align}
 H & =   \frac{2}{5} G_0 g_s \left(\re \Omega_{\rm CY} + g_s K \right) + a   d\re \left(\bar{v} \cdot \Omega_{\rm CY} \right) + \cO(g_s^{3}) \label{H3sol} \, , \\
 \label{G2sol}
 G_2 & =     d^{\dag}_{\rm CY} K  + \cO(g_s)  = - J_{\rm CY} \cdot d(4 \varphi \im \Omega_{\rm CY} - \star_{\rm CY} K) + \cO(g_s) \, , \\
G_4 & =  -\epsilon G_0 J_{\rm CY} \wedge J_{\rm CY} \left(\frac{3}{10}  + \epsilon \frac{4}{5} g_s \varphi \right)+  b J_{\rm CY} \wedge g_s^{-1} d \im v + \cO(g_s^2) \, , \\
G_6 & = 0\, ,
\end{align}
\end{subequations}   
where in the supersymmetric case
\be
\label{susypar}
\epsilon = -1 \, , \qquad a = - \oh\, , \qquad b = 1\, ,
\ee
and in the non-supersymmetric case
\be
\label{nonsusypar}
\epsilon = 1 \, , \qquad a = \frac{1}{10} \, , \qquad b =  - \frac{1}{5}\, .
\ee
Finally, $v$ is a (1,0)-form determined by
\be
v  = g_s \p_{\rm CY} f_\star + \cO(g_s^3)\, , \qquad \text{with} \qquad \Delta_{\rm CY} f_\star  = - g_s 8 G_0 \varphi \, .
\ee 

\subsubsection*{4d membranes}

In this background, one may consider branes that correspond to membranes in 4d. There are three different kinds of such membranes that are BPS objects in $\cN=1$ vacua. D8-branes wrapping the whole internal manifold $X_6$, NS5-branes wrapping special Lagrangian three-cycles of $X_6$ and D4-branes wrapping (anti)holomorphic two-cycles of $X_6$.

Let us consider a D4-brane wrapping an (anti)holomorphic two-cycle $\Sigma$ of $X_6$. Crossing such a membrane in 4d induces a change in the quanta of the internal four-form flux, scanning over the infinite family of flux vacua found in \cite{DeWolfe:2005uu}. To see if such a membrane induces a non-perturbative instability  one can dimensionally reduce the DBI+CS action of the D4-brane in the probe approximation, as done in \cite{Aharony:2008wz,Narayan:2010em}. This can be interpreted as computing the 4d membrane charge $Q$ and tension $T$, and if $Q >T$ one expects an instability similar to the one of \cite{Maldacena:1998uz}. This computation was performed in \cite{Aharony:2008wz,Narayan:2010em} for D4-branes in both cases $\epsilon = \pm 1$, in the smeared approximation. This corresponds to only consider the leading terms of the background expansion \eqref{solutionsu3} and \eqref{solutionflux}, which yield a Calabi--Yau metric, and more precisely to set $\varphi = k =0$ in those expressions. The computation was extended to the corrected backgrounds in \cite{Marchesano:2021ycx}, which can be interpreted as a one-loop correction to the DBI+CS expressions of \cite{Aharony:2008wz,Narayan:2010em}, and more precisely to the effect of a crosscap diagram between such   D4-branes and the O6-planes. At this level of accuracy it was found in \cite{Marchesano:2021ycx} that in 4d Planck units 
\be
T_{\rm D4} =  e^{K/2} \frac{1}{\ell_s^2} \left| \int_\Sigma J_{\rm CY} \right|\, , \quad Q_{D4} =   e^{K/2} \frac{\epsilon \eta}{\ell_s^2} \int_\Sigma J_{\rm CY} -  \frac{a5}{3 G_0} dd^\dag_{\rm CY} \left(f_\star  J_{\rm CY}  \right) = e^{K/2} \frac{\epsilon \eta}{\ell_s^2} \int_\Sigma J_{\rm CY}\, ,
\ee
where $\eta = {\rm sign} \, G_0$, $\eps,a$ are as in \eqref{susypar} and \eqref{nonsusypar} and $K$ is the 4d K\"ahler potential. By appropriately choosing the orientation of $\Sigma$, or equivalently by considering D4-branes or anti-D4-branes on holomorphic cycles, one can get $ Q_{\rm D4}=T_{\rm D4}$, which correspond to marginal domain walls, but not $ Q_{\rm D4}>T_{\rm D4} $. Thus, in order to check the refinement of the Weak Gravity Conjecture made in \cite{Ooguri:2016pdq} one should compute further terms in the background expansion given above.

In models with background D6-branes, that is with $N > 0$ in \eqref{tadpole2}, there is second kind of 4d membranes obtained from D-branes that are BPS in $\CN=1$ vacua. These are D8-branes wrapped on the whole of $X_6$, whose description is more involved than those of D4-branes. First, they can host harmonic (1,1) primitive worldvolume fluxes $\cF_h$, which together with the curvature corrections modify the DBI+CS action and induce D4-brane and tension. Taking these two effects into account one obtains a total tension of the form
\be
T_{\rm D8}^{\rm total} = T_{\rm D8} + \left(K^F_a - K_a^{(2)}\right)  T_{\rm D4}^a\, ,
\label{TD8tot}
\ee
with $T_{\rm D8} = e^{K/2} {\cal V}_{\rm CY}$ and   $T_{\rm D4}^a = e^{K/2} t^a$, where $J_{\rm CY} = t^a \omega_a$ defines the K\"ahler moduli. Also
\be
 K_a^{(2)} = - \frac{1}{24\ell_s^{6}} \int_{X_6} c_2(X_6) \wedge \omega_a \qquad \text{and} \qquad K^F_a = \frac{1}{2\ell_s^6} \int_{X_6} \cF_h \wedge \cF_h \wedge \omega_a\, .
\label{Kcorr}
\ee
It is important to notice that in our conventions both $ K_a^{(2)} T_{\rm D4}^a$ and $K^F_a T_{\rm D4}^a$ are non-negative quantities. In addition, one can always set $K^F_a=0$ via setting $\cF_h =0$. 

A second important feature of these D8-branes is that they have D6-branes ending on them, to cure the Freed-Witten anomaly induced by the $H$-flux \cite{Maldacena:2001xj}. In 4d terms, a membrane of this sort induces a jump in the flux quantum $m$ when crossing its worldvolume, so there should be a corresponding jump in $N$ in order to satisfy \eqref{tadpole2} at both sides of the membrane.  For a single D8-brane we have the following transition\footnote{As we will see, such D8-branes oftentimes go in pairs. However, their  jump \eqref{jump} should be considered separately.} 
\begin{equation}
    m \to m + 1 \,\implies \, q_{\a, \eta} \to  q_{\a, \eta} + \hat q_{\a, \eta} \quad \text{with} \quad \sum_{\eta} \hat q_{\a, \eta} =  h p_\a\, ,
    \label{jump}
\end{equation}
where  $\hat q_{\a, \eta} \geq 0$, and the upper bound $q_{\a, \eta} \leq 4$ should always be respected.  
At the level of accuracy with which we are describing the 10d background, this feature manifests itself as a BIon-like profile developed by the D8-brane \cite{Marchesano:2021ycx}. This profile is slightly more involved than the simplest examples \cite{Gibbons:1997xz,Evslin:2007ti}, but it contains similar features. We have a non-closed piece of the D8-brane worldvolume flux that reads\footnote{For the simplest configuration in which D6-branes are equally distributed on top of the O6-plane components before and after the jump, that is $\hat{q}_{\a,\eta} = h$, $\forall \a, \eta$, we have that $ \mathcal{F}_{\rm BIon} =  G_0^{-1} d^{\dag}_{\rm CY} K+ \cO(g_s)$, as assumed in \cite{Marchesano:2021ycx}.}
\begin{equation}
    \mathcal{F}_{\rm BIon} =  \sum_{\a, \eta}  \hat{q}_{\a, \eta} \cF_{\a, \eta}  + \cO(g_s) \, , \qquad  \cF_{\a, \eta}  = d^\dag K_{\a, \eta}\, ,
    \label{eq: worldvolume flux}
\end{equation}
We also have  a non-trivial profile for the D8-brane transverse coordinate
\be
Z = z_0 - \ell_s \sum_{\a, \eta} \hat q_{\a, \eta} \varphi_{\a, \eta}\, , \qquad z_0/\ell_s \in \mathbb{R}\, . 
\ee
This BIon-like profile also contributes to the D8-brane DBI+CS action, and therefore modifies the 4d membrane charge and tension. In terms of the latter, we have an extra term in \eqref{TD8tot} 
\be
 T_{\rm D8}^{\rm BIon} =  e^{K/2} \frac{1}{2\ell_s^6} \int_{\rm X_6} J_{\rm CY} \wedge \cF_{\rm BIon}^2 + \cO(g_s^2) \, ,
\label{Tbion}
\ee
which resembles the term $K_a^F T^a_{\rm D4}$, except that it involves a different component of the worldvolume flux. In the supersymmetric background and for a BPS D8-brane, the three corrections to $T_{\rm D8}$ also appear in the 4d membrane charge, yielding as expected that $T_{\rm D8}^{\rm total}  = Q_{\rm D8}^{\rm total}$. For the non-supersymmetric background with $\epsilon = 1$ the same D8-brane develops these corrections but with opposite charge. That is
\be
Q_{\rm D8}^{\rm total} = T_{\rm D8} - \epsilon \left(K^F_a - K_a^{(2)} \right)  T_{\rm D4}^a - \epsilon  T_{\rm D8}^{\rm BIon} \, .
\label{QD8tot}
\ee
 As a result, the excess charge for such membranes  reads
\be
 Q_{\rm D8}^{\rm total} - T_{\rm D8}^{\rm total} = \left(1+\epsilon\right) \left[K_a^{(2)} T_{\rm D4}^a - K^F_a T_{\rm D4}^a - T_{\rm D8}^{\rm BIon} \right]  \, .
\label{QTtotalnosusy}
\ee
If the term in brackets is positive for some 4d  membrane the refined WGC of \cite{Ooguri:2016pdq} is verified, signalling a non-perturbative instability of the non-supersymmetric vacuum. As mentioned before, the first term inside the bracket is always non-negative, and in fact it is positive away from the boundary of the K\"ahler cone. The second one is non-positive, but it can always be chosen to vanish by appropriate choice of worldvolume fluxes. It is thus the third one that remains to analyse, which will be the subject of the next section. For concreteness we define the quantity
\be
\Delta_{\rm D8}^{\rm Bion} \equiv - e^{K/2} \frac{1}{2\ell_s^6} \int_{\rm X_6}  J_{\rm CY} \wedge \cF^2_{\rm BIon}\, , 
\label{QTbion}
\ee
that we dub as the BIonic excess charge of the membrane. A priori this quantity is comparable to the effect of curvature corrections, and it is in fact larger for Calabi--Yau geometries near a toroidal orbifold limit. In the next sections we will analyse $\Delta_{\rm D8}^{\rm Bion}$ precisely for those geometries. Remarkably, we find a very simple expression, that suggests generalisation to arbitrary Calabi--Yau geometries of the form $\Delta_{\rm D8}^{\rm Bion} = D_a T_{\rm D4}^a$, where $D_a$ depend on discrete data.


\section{Toroidal orientifolds}
\label{s:Delta}

In this section we specify the above setup to toroidal Abelian orbifolds of the form $T^6/\IZ_N$ or $T^6/(\IZ_N \times \IZ_M)$, where the covering space is a factorisable six-torus $T^6 = (T^2)_1 \times (T^2)_2 \times (T^2)_3$ and the orbifold action respects the factorisation. As we show in the next section, for these geometries one can compute the quantity \eqref{QTbion} explicitly, obtaining a simple general expression. In the following we will summarise this expression and discuss its consequences for the stability of AdS$_4$ vacua with different D6-brane configurations.

\subsection{The BIonic excess charge}
\label{ss:BIec}

In toroidal Abelian orbifolds of the form $(T^2)_1 \times (T^2)_2 \times (T^2)_3/\Gamma$, with $\Gamma = \IZ_N$ or $\Gamma= \IZ_N \times \IZ_M$, the O6-plane content in the covering space $T^6$ is characterised by a set of factorisable three-cycles, which in homology read
\be
[\Pi_{\rm O6}] = \sum_{\alpha,\eta}  [\Pi^{\rm O6}_{\a,\eta}] = \sum_\alpha  p_\a [\Pi^{\rm O6}_\a] = \sum_\alpha p_\a  \left[(n_\a^1, m_\a^1) \times (n_\a^2, m_\a^2)\times (n_\a^3, m_\a^3)\right] \, .
\label{O6sum}
\ee
Here $\a$ runs over different homology classes in the covering space, specified by the wrapping numbers $(n_\a^i, m_\a^i) \in \IZ^2$ of each factorisable three-cycle on $(T^2)_i$. The index $\eta$ runs over different representatives in the same homology class, giving rise to a multiplicity $p_\a$. If we place the existing D6-branes on top of the O6-planes, more precisely $N_{\a,\eta}$ of them on top of $\Pi^{\rm O6}_{\a,\eta}$, the background RR two-form flux is of the form $G_2 = d^\dag_{\rm CY} K$, where
\be
K =  \sum_{\a,\eta} q_{\a, \eta} K_{\a,\eta}\, , \qquad  \ell_s^2  \Delta_{\rm CY} K_{\a,\eta} = H_\a - \delta(\Pi^{\rm O6}_{\a,\eta})\, ,
\label{Ktorus}
\ee
with $ q_{\a, \eta} = 4 - N_{\a,\eta}$ and
\be
H_\a = \ell_s^3 \left(m_\a^1 dx^1 - n_\a^1dy^1\right) \wedge \left(m_\a^2 dx^2 - n_\a^2dy^2\right) \wedge \left(m_\a^3 dx^3 - n_\a^3dy^3\right)\, ,
\ee
where $(x^i,y^i)$ are the period-one coordinates of $(T^2)_i$. From here one can extract the quantities $\varphi$ and $k$ that appear in \eqref{formK}, and describe the full background \eqref{solutionsu3} and \eqref{solutionflux}. 

Additionally, given a D8-brane-mediated flux jump of the form \eqref{jump}, the BIon-like solution that describes the D8/D6-brane system features a coexact worldvolume flux of the form  \eqref{eq: worldvolume flux}.  As a consequence we have that \eqref{QTbion} is of the form
\begin{equation}
    \Delta_{\rm D8}^{\rm Bion} =  \oh \sum_{\alpha,\beta, \eta, \zeta} \hat{q}_{\a,\eta} \hat{q}_{\b,\zeta}\, {\Delta}_{\alpha,\eta;\beta,\zeta}\, , \qquad {\Delta}_{\alpha,\eta;\beta,\zeta} = - e^{K/2} \frac{1}{\ell_s^{6}} \int_{\rm X_6}  J_{\rm CY} \wedge \cF_{\a,\eta} \wedge \cF_{\b,\zeta} .
    \label{Deltasum}
\end{equation}

From our explicit computations in the next section we moreover obtain the following results:

\begin{itemize}
    \item The integral in \eqref{Deltasum} is non-zero only when the intersection number $I_{\alpha\beta} = [\Pi^{\rm O6}_\alpha] \cdot [\Pi^{\rm O6}_\beta] = 0$ and $[\Pi^{\rm O6}_\alpha] \neq [\Pi^{\rm O6}_\beta]$, which in particular implies that $\Delta_{\alpha,\eta;\alpha,\zeta} = 0$.  In practice, this means that non-vanishing contributions  to \eqref{Deltasum} come from $\cN =2$ sectors of the compactification, that is from pairs of D6-branes wrapping three-cycles related by an $SU(2)$ rotation. In our setup, this translates into wrapping numbers $(n_\a^i, m_\a^i)$, $(n_\b^i, m_\b^i)$,  that are similar in one two-torus $(T^2)_i$ and different in the other two. We denote these pairs of three-cycles as $\cN=2$ pairs, see figure \ref{fig: separation and intersection} for examples.

\begin{figure}[htb]
\centering
\begin{subfigure}[b]{.85\textwidth}
  \centering
     \includegraphics[scale=0.35]{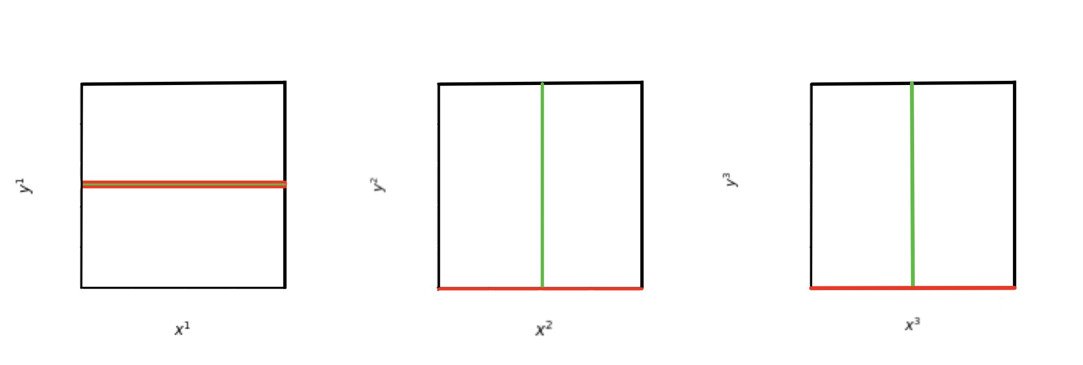}
    \caption{Diagram corresponding to an $\cN=2$ pair with one intersection over a one-cycle.}
 \end{subfigure}\\
\begin{subfigure}[b]{.85\textwidth}
    \centering
    \includegraphics[scale=0.348]{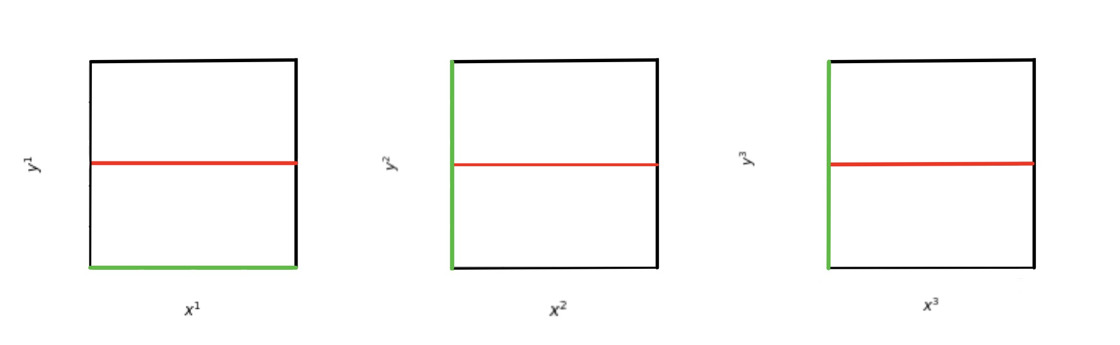}
    \caption{Diagram corresponding to an $\cN=2$ pair with  no intersection.}
    \end{subfigure}
    \caption{Configuration of 3-cycles projected over $T^2\times T^2\times T^2$ that contribute to \eqref{Deltasum} in the $T^6/\mathbb{Z}_2\times\mathbb{Z}_2$ orbifold.}
    \label{fig: separation and intersection}
\end{figure}
 \item Given a $\cN=2$ pair $(\a,\eta;\b,\zeta)$, the integral in \eqref{Deltasum} depends separately on the indices $\a,\b$ that describe the homology classes $ [\Pi^{\rm O6}_\alpha]$ and $[\Pi^{\rm O6}_\beta]$, and the indices $\eta,\zeta$ that specify the representatives.  The dependence in $\a,\b$ corresponds to the number of regions of minimal separation between $\Pi^{\rm O6}_\alpha$ and $\Pi^{\rm O6}_\beta$, which we dub $\cN=2$ subsectors. For instance, if $\Pi^{\rm O6}_\alpha$ and $\Pi^{\rm O6}_\beta$ intersect over one-cycles, the number of $\cN=2$ subsectors is the number of intersections. To measure this number we define
 \begin{equation}
    \#(\Pi_\a \cap \Pi_\b)_i = |n_\a^j m_\b^j - n_\b^jm_\a^j| \times |n_\a^k m_\b^k - n_\b^km_\a^k|\, ,
    \label{interdef}
\end{equation}
where $i \neq j \neq k$. When $\Pi^{\rm O6}_\alpha$ and $\Pi^{\rm O6}_\beta$ have parallel one-cycles in $(T^2)_i$ but they do not coincide, \eqref{interdef} does not count intersections, but instead regions of minimal separation between the two three-cycles. In both cases, \eqref{interdef} amounts to the number of `intersections' in the two two-tori where $\Pi^{\rm O6}_\alpha$ and $\Pi^{\rm O6}_\beta$ are not parallel, it is non-vanishing for a single choice of $i$, and because each $\CN=2$ subsector contributes equally to the integral in \eqref{Deltasum}, $\Delta_{\a,\eta;\b,\zeta}$ is proportional to this number. 

\item The dependence on the indices $\eta,\zeta$ arises because $\Delta_{\a,\eta;\b,\zeta}$ is different if   $\Pi^{\rm O6}_{\alpha,\eta}$ and $\Pi^{\rm O6}_{\beta,\zeta}$ intersect or not. In general, the contribution of each $\CN=2$ subsector to the integral in \eqref{Deltasum} is proportional to $t^i$, which is the area of the $(T^2)_i$ selected by \eqref{interdef}, or in other words the two-torus where $\Pi^{\rm O6}_{\alpha,\eta}$ and $\Pi^{\rm O6}_{\beta,\zeta}$ are parallel. The coefficient of the contribution depends on whether  these two three-cycles intersect or not. If they intersect over a one-cycle on $(T^2)_i$, each 
$\CN=2$ subsector contributes to the integral $-\ell_s^{-6}\int_{T^6} J_{\rm CY} \wedge \cF_{\a,\eta} \wedge \cF_{\b,\zeta}$  over the covering space  as
\begin{equation}
   \frac{t^i}{12} \,. 
   \label{intercont}
\end{equation}
If instead $\Pi^{\rm O6}_{\alpha,\eta}$ and $\Pi^{\rm O6}_{\beta,\zeta}$ do not overlap, but they are only parallel in $(T^2)_i$ we obtain\footnote{In the toroidal orientifold geometries that we consider in the next section, an $\cN=2$ pair of O6-planes that do not intersect are separated at mid-distance in their common transverse space in $(T^2)_i$. When we consider D6-branes wrapped in the same homology classes $[\Pi^{\rm O6}_{\alpha}]$ and $[\Pi^{\rm O6}_{\beta}]$ but not on top of orientifold planes in $(T^2)_i$, their BPS locations form a discretum analogous to the ones in \cite{Gomis:2005wc,Marchesano:2006ns}, because the presence of $H$-flux implies that only at certain discrete locations the D6-brane worldvolume flux $\cF = B|_{\Pi_{\rm D6}} +\frac{\ell_s^2}{2\pi}F$ can vanish. In this case, the separation between three-cycles is of the form $\frac{\ell_s^2 t_i}{L} \frac{k}{2P}$, where $L$ the length of the one-cycle wrapped in $(T^2)_i$, $P \in \mathbb{N}$ is determined by the quanta of $H$-flux, and $0\leq k \leq 2P$ is an integer. Given this separation, the contribution of this $\CN=2$ D6-brane pair to the integral $-\ell_s^{-6}\int_{T^6} J_{\rm CY} \wedge \cF_{\a,\eta} \wedge \cF_{\b,\zeta}$  is given by 
\begin{equation}
\nonumber
    \oh\left(\frac{1}{6} - \frac{k}{2P}\left(1-\frac{k}{2P}\right)\right) t^i\, ,
\end{equation}
which reduces to \eqref{intercont} for $k=0$ and to \eqref{interpar} for $k=P$.\label{OSL}}
\begin{equation}
   -\frac{t^i}{24} \,. 
   \label{interpar}
\end{equation}
Integrating over $X_6$, we divide both results by the orbifold group $\Gamma$ order,  dubbed $N_\Gamma$.

\end{itemize}

Adding all these results together, we end up with the following expression for the BIonic contribution to the 4d membrane excess charge:
\begin{equation}
     \Delta_{\rm D8}^{\rm Bion} = \frac{1}{24N_\Gamma} \sum_{(\a,\eta;\b,\zeta)\in \cN=2} \hat{q}_{\a,\eta} \hat{q}_{\b,\zeta}\, \varepsilon_{\eta\zeta} \, \#(\Pi_\a \cap \Pi_\b)_i T^i_{\rm D4}\, .
     \label{finalDelta}
\end{equation}
Here $T^i_{\rm D4} = e^{K/2} t^i$ corresponds to the 4d membrane tension of a  D4-brane wrapped around $(T^2)_i$, while $\varepsilon_{\eta\zeta}=2$ for intersecting $\cN=2$ pairs and $\varepsilon_{\eta\zeta}=-1$ for those at mid-distance. Note that in the above expression the factor of $2$ associated to the exchange of $\mathcal{F}_{\alpha,\eta}$ and $\mathcal{F}_{\beta,\zeta}$ in \eqref{Deltasum} has already been accounted for, so that we sum over each $\cN=2$ pair only once. 

Finally, we find that in general a D8-brane with a worldvolume flux is not invariant under the orientifold action, and therefore we need to consider two of them. This reflects the fact that in Calabi-Yau orientifolds oftentimes the quantum of Romans mass must be even. In fact, if we insist of working with a toroidal orbifold geometry the quantisation conditions for $m$ and other background fluxes become even more restrictive, as we now turn to discuss.

\subsection{Flux quantisation and blow-up modes}
\label{ss:fluxquant}

In the absence of localised sources the Bianchi identities \eqref{IIABI} are quite trivial, in the sense that $e^{-B} \wedge {\bf C}$ is globally well-defined. Then the quantisation condition for NS and RR fluxes read
\begin{equation}
\frac{1}{\ell_s^{p}} \int_{\Pi_{p+1}} \bar{G}_{p+1}   \in  \IZ\, , \qquad  \frac{1}{\ell_s^{2}}\int_{\Pi_3}  H \in \IZ\, .
\label{Pqu}
\end{equation} 
When we include localised sources like D-branes, we need to substitute these conditions by Page charge quantisation \cite{Marolf:2000cb}. Nevertheless, we can still make use of the quanta defined in \eqref{Pqu}, which are in fact the flux quanta used to describe the compactification in the smeared approximation. 

Additionally, the presence of O-planes can affect the quantisation of those fluxes that are not sourced by any localised object. Indeed, as pointed out in \cite{Frey:2002hf}, in type IIB orientifold compactifications that only contain O3-planes with negative charge and tension (dubbed O3$^-$) the quanta of NS and RR background three-form fluxes must be even integers. This observation was applied to toroidal orbifold geometries in \cite{Blumenhagen:2003vr,Cascales:2003zp}, where it was found that three-form flux quanta in the  covering space should be multiples of $2M$ if no flux along collapsed three-cycles was to be involved, with $M \in \IZ$ depending on the particular orbifold. 

Clearly, these type IIB orientifold constraints must have a counterpart in our type IIA setup. Let us for instance take the type IIB setup of \cite{Frey:2002hf}, with 64 O3$^-$ on a $T^6$. An NS flux of the form $H = h\, dy^1 \wedge dy^2 \wedge dy^3$ is consistent if $h \in 2\IZ$. By performing three T-dualities along $\{x^1, x^2, x^3\}$ one recovers type IIA on $T^6$ with 8 O6$^-$ that extend along such coordinates. Assuming a factorised metric, this T-duality does not affect the $H$-flux that we have considered, and so one concludes that a type IIA  $H$-flux integrated over a three-cycle that intersects an even number of O6$^-$ must be quantised in terms of even integers. The same reasoning can be applied by T-dualising the type IIB RR three-form flux along any three-cycle of $T^6$. By doing so, we recover that $G_0$, $\bar{G}_2$, $\bar{G}_4$, $\bar{G}_6$ should also correspond to even integer quanta in the said type IIA background. In general, we expect a similar statement to apply in a smooth Calabi-Yau geometry $X_6$, whenever a $p$-cycle intersects an even number of O6$^-$. 

The orbifold geometries $X_6 = T^6/\Gamma$ that we consider in the next section do contain O6$^-$, but their homology classes are more involved than that of $T^6$. The difference mostly resides in the orbifold twisted sector, which corresponds to a set of cycles that are collapsed in the orbifold limit of a smooth Calabi--Yau. Since they are collapsed, the approximation of diluted fluxes that leads to the solution \eqref{solutionsu3} and \eqref{solutionflux} is justified as long as the background fluxes do not have components on the twisted sector. Here is where the logic of \cite{Blumenhagen:2003vr,Cascales:2003zp} applies, and as a result the flux quanta computed in the covering space $T^6$ must be multiples of $2M$, for some $M \in \IZ$. In the following we will discuss how these quantisation conditions look like in the case of the $\IZ_2 \times \IZ_2$ orientifolds mirror dual to the ones considered in  \cite{Blumenhagen:2003vr,Cascales:2003zp}. 

\subsubsection*{The $\IZ_2 \times \IZ_2$ orbifold}

Let us consider a $\IZ_2 \times \IZ_2$ orbifold over the factorisable six-torus $T^6= (T^2)_1 \times (T^2)_2 \times (T^2)_3$. The complex coordinate describing each two-torus is given by
\begin{equation}
    z^i=2\pi R_i(x^i+ i u_i y^i)\, ,
    \label{coordZ2Z2}
\end{equation}
with $x^i$ and $y^i$ real coordinates of unit periodicity, $u_i \in \mathbb{R}$ describing the complex structure and $t^i = 4\pi^2 \ell_s^{-2} R_i^2 u_i$ the K\"ahler moduli of each $T^2$. The generators of the orbifold group act as
\be
\theta\, :\, (z^1,z^2,z^3) \mapsto (-z^1,-z^2,z^3)\, , \qquad \omega\, :\, (z^1,z^2,z^3) \mapsto (z^1,-z^2,-z^3)\, ,
\label{eq: z2z2 action}
\ee
leaving fixed the coordinate values $x^i, y^i = \{0, 1/2\}$. Such coordinates correspond to the orbifold twisted sector, which can be interpreted as a set of collapsed cycles. The nature of these cycles depends on the choice of discrete torsion \cite{Font:1988mk,Vafa:1994rv}, which specifies how $\omega$ acts on the fixed point set of $\theta$, and so on. With one choice of discrete torsion the twisted sector corresponds to 48 collapsed two-cycles and 48 collapsed four-cycles, and the orbifold cohomology amounts to $(h^{1,1}, h^{2,1})_{\rm orb} = (51,3)$, while for the second choice it correspond to 96 collapsed three-cycles and  $(h^{1,1}, h^{2,1})_{\rm orb} = (3,51)$. These two choices are related to each other by mirror symmetry. 

We can now apply the orientifold quotient $\Omega_p (-1)^{F_L}{\cal R}$, with
\be
{\cal R} \, :\, (z^1,z^2,z^3) \mapsto (\bar{z}^1,\bar{z}^2,\bar{z}^3)\, .
\ee
This generates four different kinds of O6-planes:
\begin{subequations}	
	\label{O6Z2Z2}
\begin{align}
[\Pi^{\rm O6}_{\cal R}] & = \left[(1, 0) \times (1, 0) \times (1, 0)\right]\, , \\
[\Pi^{\rm O6}_{{\cal R}\theta}] & = \left[(0, 1) \times (0, -1)\times (1, 0)\right]\, , \\
[\Pi^{\rm O6}_{{\cal R}\omega}] & = \left[(1, 0) \times (0, 1) \times (0,-1)\right]\, , \\
[\Pi^{\rm O6}_{{\cal R}\theta\omega}] & = \left[(0, -1) \times (1,0) \times (0,1)\right]\, ,
\end{align}
\end{subequations}   
each labelled by the orientifold group element that leaves them fixed. The multiplicity of each O6-plane class is $p_\a =8$, and they go over the different orbifold fixed points, so the index $\eta$ is better represented by the vector $\vec\eta = (\eta_1, \eta_2,\eta_3)$ with $\eta_i = 0, 1/2$. While the fixed loci are the same, the O6-plane nature is different for both choices of discrete torsion. For $(h^{1,1}, h^{2,1})_{\rm orb} = (51,3)$ all of them are O6$^-$, while for $(h^{1,1}, h^{2,1})_{\rm orb} = (3,51)$ one of the four classes in \eqref{O6Z2Z2} has to correspond to O6$^+$-planes \cite{Angelantonj:1999ms}. Thus, in this second case, by placing D6-branes on top of the O6-planes one will never be able to construct a model absent of NS tadpoles, even in the presence of fluxes.\footnote{One could do so by introducing D6-branes at angles \cite{Marchesano:2004xz,Blumenhagen:2005tn}, but these more involved configurations will not be considered here.} For this reason in the following we will focus on the case where $(h^{1,1}, h^{2,1})_{\rm orb} = (51,3)$. 

Let us now see what is the appropriate flux quantisation in the  $\IZ_2 \times \IZ_2$ orientifold with $(h^{1,1}, h^{2,1})_{\rm orb} = (51,3)$. In the absence of orientifold projection one can use the results of \cite{Blumenhagen:2003vr}, that show that the integral lattice of three-cycles is of the form $2[\Pi_\a]$, where $[\Pi_\a] = \left[(n_\a^1, m_\a^1) \times (n_\a^2, m_\a^2)\times (n_\a^3, m_\a^3)\right]$ is an integer three-cycle in the covering space $T^6$. If we now apply our criterium for flux quantisation in the presence of O6$^-$-planes we obtain that the $H$-flux must be quantised in units of 4 from the viewpoint of $T^6$. That is, $[\ell_s^{-2} H] = \sum_\a 4 h_\a  {\rm P.D.} [\Pi_\a]$, with $h_\a \in \IZ$. In particular, if as before we consider a flux of the form  $[\ell_s^{-2} H] =  h  {\rm P.D.} [\Pi_{\rm O6}]$, we find that $h \in \IZ/2$.  

This quantisation in units of four is quite reminiscent of a similar condition for D6-branes. Indeed, for this choice of discrete torsion the minimal amount of covering-space three-cycles needed to build a consistent boundary state is two \cite{Douglas:1998xa,Gomis:2000ej}. Then, when introducing the orientifold projection and placing the D6-branes on top of an O6-plane one finds that its gauge group is $USp(2N)$, which means that each D6-brane in the orientifolded theory corresponds to four D6-branes in the covering space \cite{Cvetic:2001nr}. In other words, the charges $q_{\a,\eta}$ that appear in \eqref{O6D6source} are quantised in units of 4. 

Let us finally turn to the quantisation of internal RR fluxes. In this case one can directly use the results of \cite{Cascales:2003zp} on a type IIB mirror symmetric orientifold, because both the RR fluxes and the D-branes that generate them have a simple behaviour under T-duality. It was found in  \cite{Cascales:2003zp} that covering-space RR three-form fluxes must be quantised in units of 8 if one does not want to turn them on along twisted three-cycles. In our type IIA setup, this means that the quanta of Romans mass $m$ and that of four-form flux must also be quantised in units of 8 if one wants to maintain the orbifold geometry $T^6/(\IZ_2 \times \IZ_2)$. From the type IIA perspective the quantisation in units of 8 of the Romans mass may seem surprising, but one can understand it in terms of the D-brane object that generates $G_0 =\ell_s^{-1}m$, namely a D8-brane wrapped on the internal space. Such a D8-brane will have induced D4-brane charge in the twisted sector, due to the curvature corrections and the non-trivial B-field at the orbifold point. The results of \cite{Blumenhagen:2003vr,Cascales:2003zp}  imply that, in order to construct a D8-brane boundary state with no induced twisted charges, one needs four of them in the covering space to form the regular representation of the orbifold group. The orientifold then doubles this number to eight D8-branes. In terms of fluxes, if one wants to have a non-vanishing Romans mass without inducing any four-form flux on the orientifold twisted sector one must impose that $m$ is a multiple of 8. 

Notice that these flux quantisation conditions are quite constraining when imposing the tadpole equation \eqref{tadpole2}, as they only allow for the solution
\be
m = 8\, , \qquad h = \oh\, , \qquad N=0\, ,
\label{orbisolBI}
\ee
which contains no D6-branes at all. Thus, a domain-wall transition of the form \eqref{jump} is not allowed starting from this orientifold vacuum, because the quantum of Romans cannot be any larger, and this applies to both supersymmetric and non-supersymmetric vacua. 

Nevertheless, one can apply the same philosophy of \cite{DeWolfe:2005uu} and consider orientifold vacua in which the K\"ahler moduli of the twisted sector have been blown up due to the presence of a four-form flux along them, see Appendix \ref{ap:Z2xZ2}. In this case we no longer need to impose that $m$ is a multiple of 8, but only impose the orientifold constraint that sets it as an even integer. Therefore we have a richer set of solutions to the tadpole constraint \eqref{tadpole2}, like the family 
\be
m = 2k\, , \qquad h = \oh\, , \qquad N=4-k\, ,\qquad k = 1,2,3,4\, ,
\label{blowsolBI}
\ee
or
\be
m = 2k\, , \qquad h = 1 \, , \qquad N=4-2k\, ,\qquad k = 1,2\, .
\label{blowsolBI2}
\ee
Moreover, if as in \cite{DeWolfe:2005uu} we make a choice of four-form flux such that the blow-up K\"ahler moduli are much smaller than the toroidal ones, then the result \eqref{finalDelta} should be a good approximation for the BIonic D8-brane excess-charge in $\cN=0$ vacua. Indeed, when twisted K\"ahler moduli are blown up both $J_{\rm CY}$ and ${\cal F}$ will be modified and so will be $\Delta_{\rm D8}^{\rm Bion}$, but one expects an effect that is of the order of the size of the blown-up two-cycles. Therefore, if we blow up the twisted two-cycles but their size remains much smaller than the toroidal K\"ahler moduli, we expect \eqref{finalDelta} to give us a good approximation of the BIonic D8-brane excess charge. 

Given the value of $\Delta_{\rm D8}^{\rm Bion}$, one should finally compare it with $\Delta_{\rm D8}^{\rm curv} \equiv K_a^{(2)} T^a_{\rm D4}$, which one can again compute in the orbifold limit. For this computation the relevant intersection number is $c_2(X_6) . R_i$, where $R_i$ is the sliding divisor defined in Appendix \ref{ap:Z2xZ2}. Using the results of \cite{Denef:2005mm} one finds that $c_2(X_6) . R_i = 24$ and therefore
\be
\Delta_{\rm D8}^{\rm curv} = \oh  \left(T^1_{\rm D4} + T^2_{\rm D4} + T^3_{\rm D4}\right)\, ,
\label{D8curvz2xz2}
\ee
see Appendix \ref{ap:Z2xZ2curv} for details. Recall that in order to satisfy the refined WGC for 4d membranes, it should be that $\Delta_{\rm D8}^{\rm curv} + \Delta_{\rm D8}^{\rm Bion} > 0$ for any BIon configuration.

\subsection{BIon configurations and the WGC}

One can check that \eqref{finalDelta} reproduces the result  obtained in \cite{Marchesano:2021ycx} for the orientifold $T^6/(\Z_2 \times \IZ_2)$ and a transition \eqref{jump}  in which $\hat{q}_{\a,\eta} = h$, $\forall \a, \eta$.  Indeed, there are six different pairs of different homology classes. For each combination there are 64 $\cN=2$ pairs, 32 of which intersect and 32 which do not, and each of them with a single $\cN=2$ subsector. The parallel one-cycles correspond to the basis $\{[(1,0)^i], [(0,1)^i]\}$, $i=1,2,3$ of $H_1(T^6, \IZ)$, and so each $T^i_{\rm D4}$ is selected twice in the sum \eqref{finalDelta}. Applying all these data we obtain
\begin{equation}
     \Delta_{\rm D8}^{\rm Bion}(T^6/(\IZ_2\times\IZ_2)) = \frac{h^2}{24 \cdot 4} 32  (2-1) 2 \left(T^1_{\rm D4} + T^2_{\rm D4} + T^3_{\rm D4}\right)\, ,
     \label{DeltaZ2xZ2}
\end{equation}
and so \cite[eq.(7.29)]{Marchesano:2021ycx} is recovered. However, such a transition is never realised as a jump between AdS$_4$ vacua. Indeed, we have seen that $q_{\a,\eta}$ must be a multiple of 4, so if we have the same number of D6-branes on top of each orientifold it means that the negative charge and tension of each O6$^-$-plane is cancelled, and necessarily $mh=0$ in \eqref{tadpole2}. In other words, we are in a 4d Minkowski vacuum. The second option for this equal distribution of D6-branes is to have none at all, which takes us back to an AdS$_4$ vacuum in which $mh=4$, like the one in \eqref{orbisolBI}. A transition between these two $\Z_2 \times \IZ_2$ orientifold vacua is not mediated by 4d membrane arising from a BIonic D8-brane, but instead from a bound state of D8-brane, D4-brane and NS5-brane. The 4d vacuum of larger energy is $\cN=1$ Minkowski, and the membrane bound state is BPS and satisfies a no-force condition regardless of whether we jump to a $\cN=1$ or $\cN=0$ AdS$_4$ vacuum, as expected from the general results of \cite{Herraez:2020tih,Lanza:2020qmt}. 

Transitions mediated by a BIonic D8-brane for instance arise when increasing the value of $k$ in the family of vacua \eqref{blowsolBI} and \eqref{blowsolBI2} which, as explained, take us away from the orbifold limit. If we are in a non-supersymmetric vacuum of the sort discussed in section \ref{s:nonsusy}, the BIon excess charge should be computed to a good approximation by \eqref{finalDelta}, which will depend on how the D6-branes are arranged before and after the jump. In general we will have $8(4-2kh)$ D6-branes distributed in groups of 4 on the three-cycles $\Pi^{\rm O6}_{\a,\eta}$ within each homology class in \eqref{O6Z2Z2}. 

\begin{figure}[h]
\centering
\begin{subfigure}[b]{.415\textwidth}
  \centering
    \hspace{-0.32cm} \includegraphics[width=\textwidth]{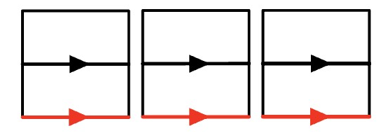}
    \caption*{$\Pi^{\rm O6}_{\cal R}  $}
 \end{subfigure}\hfill
\begin{subfigure}[b]{.415\textwidth}
  \centering
 \includegraphics[width=\textwidth]{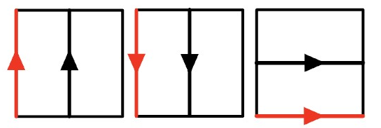}    \hspace{-0.36cm}
    \caption*{$\Pi^{\rm O6}_{{\cal R}\theta} $}
\end{subfigure}\\
\begin{subfigure}[b]{.4\textwidth}
    \centering
    \includegraphics[width=\textwidth]{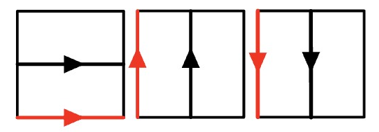}
    \caption*{$\Pi^{\rm O6}_{{\cal R}\omega}  $}
\end{subfigure}\hfill
\begin{subfigure}[b]{.4\textwidth}
    \centering
    \includegraphics[width=\textwidth]{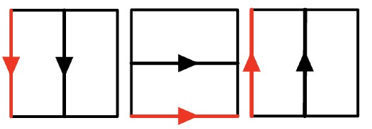}
    \caption*{$\Pi^{\rm O6}_{{\cal R}\theta\omega} $}
    \end{subfigure}
\caption{D6-brane configuration leading to \eqref{eq: intersecting D6branes example}. In red are the O6-planes with D6-branes on top of them.}
\label{fig: z2z2 intersecciones}
\end{figure}

For simplicity, we may consider the case where for each value of $\a$ all $8(4-2kh)$ D6-branes are on a single three-cycle, that is in a given choice of $\eta$. For instance, one may consider the case that such D6-branes are on top of the four O6-planes that go through the origin, which corresponds to selecting $\vec{\eta} = (0,0,0)$ for each value of $\a$, as represented in figure \ref{fig: z2z2 intersecciones}. Then one can apply \eqref{finalDelta} to compute the BIon excess charge of a single D8-brane, without taking into account its orientifold image. In this case we have that 
\be
\hat{q}_{\a, (0,0,0)} = 8h\, , \ \hat{q}_{\a, \vec{\eta} \neq (0,0,0)} = 0\quad  \forall \a\, , \qquad \varepsilon_{(0,0,0),(0,0,0)} = 2 \, , \qquad  \#(\Pi_\a \cap \Pi_\b) = 1 \, ,
\label{eq: intersecting D6branes example}
\ee
and that each two-torus is selected twice by the pairwise intersection. Therefore
\begin{equation}
     \Delta_{\rm D8}^{\rm Bion} = \frac{8h^2}{3} \left(T^1_{\rm D4} + T^2_{\rm D4} + T^3_{\rm D4}\right)\, ,
     \label{DeltaZ2xZ2int}
\end{equation}

\begin{figure}[h]
\centering
\begin{subfigure}[h]{.4\textwidth}
  \centering
    \includegraphics[width=\textwidth]{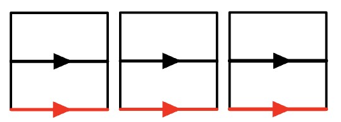}
    \caption*{$\Pi^{\rm O6}_{\cal R}$}
 \end{subfigure}\hfill
\begin{subfigure}[h]{.4\textwidth}
  \centering
 \includegraphics[width=\textwidth]{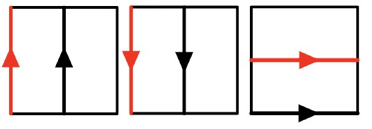}   
    \caption*{$\Pi^{\rm O6}_{{\cal R}\theta}$}
\end{subfigure}\\
\begin{subfigure}[h]{.4\textwidth}
    \centering
    \includegraphics[width=\textwidth]{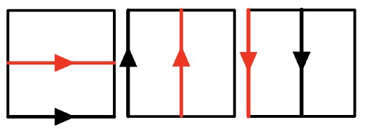}
    \caption*{$\Pi^{\rm O6}_{{\cal R}\omega}  $}
\end{subfigure}\hfill
\begin{subfigure}[h]{.4\textwidth}
    \centering
    \includegraphics[width=\textwidth]{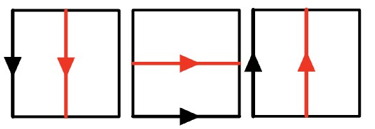}
    \caption*{$\Pi^{\rm O6}_{{\cal R}\theta\omega} $}
    \end{subfigure}
    \caption{D6-brane configuration that leads to \eqref{eq: non intersecting D6branes example}. In red are the O6-planes with D6-branes on top of them.}
    \label{fig: z2z2 no intersecciones}
\end{figure}
\noindent
signalling an instability of the vacuum. One can also consider a configuration in which the D6-branes do not intersect among each other, like for instance in figure \ref{fig: z2z2 no intersecciones}. Then
\be
\hat{q}_{{\cal R}, (0,0,0)} =  \hat{q}_{{\cal R}\theta, (0,0,\oh)} =   \hat{q}_{{\cal R}\omega, (\oh,\oh,0)} =  \hat{q}_{{\cal R}\theta\omega, (\oh,\oh,\oh)} = 8h\, 
\label{eq: non intersecting D6branes example}
\ee
with all other $\hat{q}_{\a,\vec\eta}$ vanishing. Because there is no pair of BIon sources that intersect, $\varepsilon_{\vec\eta, \vec\zeta}=-1$ and the contributions to \eqref{finalDelta} are all negative, and more precisely we recover
\begin{equation}
     \Delta_{\rm D8}^{\rm Bion} = - \frac{4h^2}{3} \left(T^1_{\rm D4} + T^2_{\rm D4} + T^3_{\rm D4}\right)\, .
      \label{DeltaZ2xZ2noint}
\end{equation}
Taking into account the curvature correction effect \eqref{D8curvz2xz2}, one concludes that, for $h=1$, $\Delta_{\rm D8}^{\rm curv} + \Delta_{\rm D8}^{\rm Bion} < 0 $.  In this case there is an excess tension for the corresponding 4d membrane, which then does not satisfies the inequality of the Weak Gravity Conjecture. As far as D8/D6-systems are concerned, such a 4d non-supersymmetric vacuum seems non-perturbatively stable.

\subsubsection*{Caveats}

The result $\Delta_{\rm D8}^{\rm curv} + \Delta_{\rm D8}^{\rm Bion} < 0$ is surprising from the viewpoint of the WGC for 4d membranes. Indeed,  the set of $\cN=0$ vacua corresponding to \eqref{eq: intersecting D6branes example} and \eqref{eq: non intersecting D6branes example} have several independent decay channels. One consists of decreasing the four-form flux quanta via nucleation of D4-branes on two-cycles. A second one is to increase $k$ in \eqref{blowsolBI} or \eqref{blowsolBI2}, mediated by BIonic D8-branes. A third one would be to leave $m$ fixed and increase the $H$-flux quantum $h$ whenever the tadpole conditions permits, mediated by an NS5-brane wrapping a special Lagrangian three-cycle. Out of these three possibilities, only the first one is available when $k$ takes it maximal value in \eqref{blowsolBI} or \eqref{blowsolBI2}. In that case from the intuition developed in  \cite{Ooguri:2016pdq} one would expect that at least some D4-brane nucleation is favoured, leading to a non-perturbative instability. If that is the case, all vacua of this sort, including those with space-time filling D6-branes, are likely to be unstable via D4-brane nucleation, and so the AdS Instability Conjecture would be verified for this setup. As mentioned before, at this level of approximation $Q_{\rm D4} = T_{\rm D4}$, and it remains as an open problem to see whether or not  $Q_{\rm D4} > T_{\rm D4}$ after further corrections are taken into account.

Whenever we have several possible decay channels involving independent 4d membrane charges, we would expect that several 4d membranes satisfy the refined WGC $Q >T$, or more precisely a Convex Hull Condition \cite{Cheung:2014vva} adapted to 4d membranes. For the vacua of the sort \eqref{eq: intersecting D6branes example} and \eqref{eq: non intersecting D6branes example} this includes at least one 4d membrane with D8-brane charge. However for $h=1$ in \eqref{blowsolBI2} we find that depending on the D6-brane positions we have either $Q_{\rm D8}^{\rm total} > T_{\rm D8}^{\rm total}$ or  $Q_{\rm D8}^{\rm total} < T_{\rm D8}^{\rm total}$. This contradicts  our WGC-based expectations, because in both cases the transition is very similar energetically. Indeed, the vacuum energy at tree level reads 
\be
V|_{\rm vac} = - \frac{16\pi}{75} e^K \cK^2 m^2 \simeq - \frac{243\pi}{50} \sqrt{\frac{3}{5}} \frac{\kappa^{3/2} h^4 |m|^{5/2}}{|\hat{e}_1\hat{e}_2 \hat{e}_3|^{3/2}}\, ,
\label{vacuumen}
\ee
where $\hat{e}_i$ are defined as in \eqref{Kahler} and correspond to the flux combinations that fix the untwisted K\"ahler moduli (that have triple intersection number  ${\cal K}_{123} = \kappa = 2$), and in the second equality we have neglected the contribution coming from blown-up two-cycles. A jump of the form $k \to k+1$ in \eqref{blowsolBI} or \eqref{blowsolBI2} not only translates into a change in $m$ but also in $\hat{e}_i$, which are negative numbers for $\cN=0$ vacua with $m>0$, see Appendix  \ref{ap:Z2xZ2}. Given that $\Delta_{\rm D8}^{\rm Bion} = D_i T_{\rm D4}^i$, it seems reasonable to assume that the full flux jump is given by
\be
m \to m + 2\, , \qquad \hat{e}_i \to \hat{e}_i +  2 K_i^{(2)} + 2 D_i \simeq  \hat{e}_i  + 1 + 2 D_i\, ,
\label{cavejump}
\ee
where for simplicity we have set $m^a=0$ in \eqref{Kahler}, and again neglected fluxes along twisted cycles. Or results above imply that $D_i = 8h^2/3$ for \eqref{eq: intersecting D6branes example} and $D_i = -4h^2/3$ for \eqref{eq: non intersecting D6branes example}, so in all cases $|\hat{e}_i|$ decreases except when $\Delta_{\rm D8}^{\rm curv} + \Delta_{\rm D8}^{\rm Bion} < 0$. While this effect increases the vacuum energy in such a case, for large values of $|\hat{e}_i|$ it is a subleading effect with respect to the increase in $|m|$. So we always decrease the vacuum energy when we perform the jump $k \to k+1$, and so there is a priori no reason why in one vacuum D8-brane nucleation is favoured and not in the other. 

In light of these considerations, let us discuss some possible loopholes in our derivation of \eqref{DeltaZ2xZ2noint}, or in its interpretation as a violation of the WGC for 4d membranes:

\begin{itemize}

\item As mentioned above, the results \eqref{DeltaZ2xZ2int} and \eqref{DeltaZ2xZ2noint} are approximations, because they are computed in terms of an integral in the orbifold covering space $T^6$. However, in order to have a transition that increases $k$ in \eqref{blowsolBI} it is necessary to consider Calabi--Yau geometries in which the twisted cycles have been blown up. This will modify the integral that leads to the general result \eqref{finalDelta}, but one expects the correction to be suppressed as the quotient $t^{\rm tw}/t^{\rm untw}$, between the typical size of a blown-up two-cycle $t^{\rm tw}$ and that of an untwisted two-cycle $t^{\rm untw}$. As follows from the analysis of Appendix \ref{ap:Z2xZ2}, this quotient can be arbitrarily small, and so it is consistent to neglect the corresponding correction to $\Delta_{\rm D8}^{\rm Bion}$. Similarly, as we blow up the twisted cycles, the excess charge \eqref{QTtotalnosusy} will receive a different contribution from the term $K_a^{(2)} T_{\rm D4}^a$, as it follows from eq.\eqref{c2J}. Again, this correction should be suppressed as $t^{\rm tw}/t^{\rm untw}$ compared to \eqref{finalDelta}, and can be neglected in the same way that they were neglected in \eqref{vacuumen}. In particular, it is highly unlikely that any of these corrections will flip the sign of  $\Delta_{\rm D8}^{\rm curv} + \Delta_{\rm D8}^{\rm Bion}$ computed in the orbifold limit.

\item The 10d supergravity solution \eqref{solutionsu3} and \eqref{solutionflux} is a perturbative expansion that fails near the O6-planes, and this could affect significantly the D8-brane BIon solution. Corrections to the integrals in \eqref{Deltasum} could come from such regions, which we treat via the regularisation scheme used in \cite[Appendix E]{Marchesano:2021ycx} and in the next section. This however seems unlikely in the examples at hand, because for  the regions in which the BIon solution blows up and needs to be regularised are those in which the D6-brane charge cancels the O6-plane negative charge or even flips it, and the 10d background is at weak coupling and well-behaved.\footnote{Notice that in addition the D6-brane configuration \eqref{eq: non intersecting D6branes example}, which is the problematic one for the WGC, displays no intersecting sources, and so it is more reliable with respect to the computation of $\Delta_{\rm D8}^{\rm Bion}$.} 

\item Assuming that the sign in \eqref{DeltaZ2xZ2noint} is correct, there could be other D8-brane that mediates a decay and has $\Delta_{\rm D8}^{\rm Bion} > 0$. For instance one could consider a  BIon profile different from \cite[eq.(7.3)]{Marchesano:2021ycx}, with lower tension. It would however be problematic if such a BIon solution existed as it would mean that, in a supersymmetric setup one would find a D8-brane with the same charges and lower tension than a BPS object. 

\item It could be that a more complicated bound state 4d membrane charges mediates the decay. Adding harmonic worldvolume fluxes to the D8-brane would not help, as this would switch on $K_a^F$ in \eqref{QTtotalnosusy} and render $Q_{\rm D8}^{\rm total} - T_{\rm D8}^{\rm total}$ even more negative. A different option is to involve NS5-branes.  It follows from \eqref{vacuumen} that in order to decrease the energy we need to increase the $H$-flux quantum $h$, which is not always an option. Indeed, if we increase $k=1 \to 2$ in \eqref{blowsolBI2} there is room to also increase $h$ without violating the tadpole condition. 

\item The expression for the vacuum energy \eqref{vacuumen} is a tree-level result, and it is subject to one-loop corrections. In particular there will be corrections coming from open string states stretching between different D6-branes. The masses of these objects are the main difference between the two configurations \eqref{eq: intersecting D6branes example} and \eqref{eq: non intersecting D6branes example}. In the first case they include light modes that will appear in the effective theory, while in the second case they are all massive modes above the compactification scale that need to be integrated out. The resulting threshold corrections will therefore be different and this could imply a change in the vacuum energy such that the decay is no longer energetically favoured in the second case. While this is an exciting possibility, it could also be that such threshold corrections to the vacuum energy are captured by the different values of $D_i$ in \eqref{cavejump}. In that case for large values of $\hat{e}_i$ the effect on the vacuum energy would be significantly suppressed and nothing would change. 

\item Finally, 4d membranes made up from D8-branes belong to the set of EFT membranes defined in \cite{Lanza:2020qmt} (see also \cite{Lanza:2019xxg}), and so their domain wall solutions can be described in 4d EFT terms. However such solutions are a priori not captured by the thin wall approximation. It could then be that because of the significant variation of the scalar fields, the criterium $Q_{\rm D8}^{\rm total} > T_{\rm D8}^{\rm total}$ is not the appropriate one to detect a non-perturbative instability. Nevertheless, if the expression $\Delta_{\rm D8}^{\rm Bion} = D_i T_{\rm D4}^i$ does translate into the flux jump \eqref{cavejump} when crossing the 4d membrane, one could apply the reasoning of \cite[Section 5]{Marchesano:2021ycx} and conclude that when  $Q_{\rm D8}^{\rm total} < T_{\rm D8}^{\rm total}$ there is no membrane nucleation. 

\end{itemize}

\section{Examples}
\label{s:examples}

In this section we present several examples of toroidal orbifolds, that illustrate how the different elements of formula describing the BIonic excess charge work together to provide the final result. We mainly focus on the $\IZ_2 \times \IZ_2$, $\mathbb{Z}_4$ and $\mathbb{Z}_3\times \mathbb{Z}_3$ orbifold groups, for which we perform the computations explicitly. We also consider, more schematically, the $\mathbb{Z}_6$ and $\mathbb{Z}_2\times \mathbb{Z}_4$ orbifolds.

In order to compute the integral $\int_{X} \mathcal{F}\wedge \mathcal{F}\wedge J$ we need to find an explicit expression for  the world-volume flux. As a first step we identify the different O6-planes and perform a Fourier expansion of the bump $\delta$-forms that describe them. The motivation for this being that the world-volume flux is determined by a set of 3-form currents $K_{\a,\eta}$ as in  \eqref{eq: worldvolume flux}, and such 3-form currents are defined through the Laplace equation \eqref{Ktorus}. Therefore, to find concrete expressions for $K_{\a,\eta}$ we need to build currents whose Laplacian returns  bump $\delta$-forms. Expanding in Fourier modes will prove to be an extremely useful tool to make this construction while controlling at the same time the connection with the smeared limit of our solution. Once these aspects are known, it is immediate to compute $\mathcal{F}_{\alpha,\eta}$ and evaluate the BIonic corrections using \eqref{Deltasum}.

\subsection{$T^6/\mathbb{Z}_2\times\mathbb{Z}_2$}

We start by revisiting in greater detail the orbifold discussed in the previous section, that is a $\mathbb{Z}_2\times\mathbb{Z}_2$ orbifold with periodic coordinates given by \eqref{coordZ2Z2} and orbifold action acting as \eqref{eq: z2z2 action}. The metric and the Kähler form are
\begin{align}
    g =&\, 4\pi^2 \ell_s^2\, {\rm diag} \left(\hat{R}_1^2, \hat{R}_2^2, \hat{R}_3^2, u_1^2\hat{R}_1^2, u_2^2\hat{R}_2^2, u_3^2\hat{R}_3^2 \right)  \, ,
    \\
    J=& \, \ell_s^2( t^1 dx^1\wedge dy^1 +t^2 dx^2\wedge dy^2 +t^3 dx^3\wedge dy^3). \label{eq: kahler form z2z2}
\end{align}
where we have defined the dimensionless radii $\hat{R}_i=R_i/\ell_s$ and the K\"ahler moduli $t^i=4\pi^2 \hat{R}_{i}^2 u_i$.

It is worth noting that the choice of complex structure  \eqref{coordZ2Z2} is not the only one compatible with the $\mathbb{Z}_2\times \mathbb{Z}_2$ symmetry. For each of the two-tori we are free to choose the complex structure as $\tau^i=iu_i$ or $\tau^i=1/2+iu_i$. From this point onward we will focus on the case where all the tori follow the former choice, as in \eqref{coordZ2Z2}. Results are similar for the other possible choices. 

The orientifold planes are given by the fixed points of the orientifold involution $\sigma(z)=\bar{z}$, up to orbifold action identifications. Consequently, we have the four different kinds of orientifold planes, summarised in table \ref{table: z2z2 orientifolds} and already introduced in \eqref{O6Z2Z2}. They are schematically represented as the arrow segments (both red and black) in figures \ref{fig: z2z2 intersecciones} and \ref{fig: z2z2 no intersecciones}.

\renewcommand{\arraystretch}{0.9}
\begin{table}[H]
$$
\begin{array}{|l|l|c|}
\hline \Pi_{\alpha} & \text { Fixed point equation } & \text { O6-plane position }  \\
\hline \Pi_{0} & \sigma\left(z^{a}\right)=z^{a} & y^{1} \in\left\{0,\frac{1}{2}\right\} \quad y^{2} \in\left\{0,\frac{1}{2}\right\} \quad y^{3} \in\left\{0,\frac{1}{2}\right\} \\
\Pi_{1} & \sigma\left(z^{a}\right)=\theta\left(z^{a}\right) & x^{1} \in\left\{0,\frac{1}{2}\right\} \quad x^{2} \in\left\{0,\frac{1}{2}\right\} \quad y^{3} \in\left\{0,\frac{1}{2}\right\} \\
\Pi_{\mathcal{R}\omega} & \sigma\left(z^{a}\right)=\omega\left(z^{a}\right) & y^{1} \in\left\{0,\frac{1}{2}\right\} \quad x^{2} \in\left\{0,\frac{1}{2}\right\} \quad x^{3} \in\left\{0,\frac{1}{2}\right\} \\
\Pi_{\mathcal{R}\theta\omega} & \sigma\left(z^{a}\right)= \theta\omega\left(z^{a}\right) & x^{1} \in\left\{0,\frac{1}{2}\right\} \quad y^{2} \in\left\{0,\frac{1}{2}\right\} \quad x^{3} \in\left\{0,\frac{1}{2}\right\} \\
\hline
\end{array}
$$
\caption{O6-planes in $T^6/\mathbb{Z}_2\times \mathbb{Z}_2$.}
\label{table: z2z2 orientifolds}
\end{table}
The above content of O6-planes can be expressed in terms of invariant bulk three-cycles. This is quite simple for the current case, but it will become more nuanced in the following examples. Let $\pi_{2i-1}$ and $\pi_{2i}$ constitute a basis of fundamental one-cycles on the torus $(T^2)_i$ $(i=1,2,3)$, i.e. one-cycles winded once around the directions used for the periodic identifications that parametrized the torus in \eqref{coordZ2Z2}. Then we define the following set of toroidal three-cycles:
\be
\pi_{IJK}=\pi_I \otimes \pi_J \otimes \pi_K\, .
\ee
with $I=1,2$, $J=3,4$ and $K=5,6$. From \cite{Blumenhagen:2003vr} we know that the smallest integer toroidal cycles are
\begin{equation}
\begin{array}{ll}
\rho_{1} \equiv 2\pi_{135}, & \rho_{2} \equiv 2\pi_{136}\,,\\
\rho_{3} \equiv 2\pi_{145}, & \rho_{4} \equiv 2\pi_{146}\,,\\
\rho_{5} \equiv 2\pi_{235}, & \rho_{6} \equiv 2\pi_{236}\,,\\
\rho_{7} \equiv 2\pi_{245}, & \rho_{8} \equiv 2\pi_{246}\,.\\
\end{array} \label{eq: bulk cycles z2}
\end{equation}
Then, the orientifold plane content can be expressed in terms of these invariant cycles as %
\begin{equation}
    \Pi_{\rm O6}=4\rho_1-4\rho_7-4\rho_4-4\rho_6\,.
\end{equation}
The next step will be to construct the $\delta$-like bump functions living in the factorised orbifold structure. Taking the O6-plane positions from  Table \ref{table: z2z2 orientifolds} a delta bump function can be expressed as a product of  conventional Fourier expansions for each $T^2_{i}$ with support on the fixed loci $\Pi_{\alpha}$.

\begin{subequations}
\begin{align}
    \delta(\Pi_\mathcal{R})=&\ell_s^3\sum_{\vec{\eta}}\left[\sum_{n_1\in\mathbb{Z}} e^{2\pi i n_1 (y^1-\eta_1)} dy^1 \right]\wedge \left[\sum_{n_2\in\mathbb{Z}} e^{2\pi i n_2 (y^2-\eta_2)} dy^2 \right]\wedge \left[\sum_{n_3\in\mathbb{Z}} e^{2\pi i n_3 (y^3-\eta_3)} dy^3 \right]\, ,\\
    \delta(\Pi_{1})=&\ell_s^3\sum_{\vec{\eta}}\left[\sum_{n_1\in\mathbb{Z}} e^{2\pi i n_1 (x^1-\eta_1)} dx^1 \right]\wedge \left[\sum_{n_2\in\mathbb{Z}} e^{2\pi i n_2 (x^2-\eta_2)} dx^2 \right]\wedge \left[\sum_{n_3\in\mathbb{Z}} e^{2\pi i n_3 (y^3-\eta_3)} dy^3 \right]\, ,\\
    \delta(\Pi_{\mathcal{R}\omega})=&\ell_s^3\sum_{\vec{\eta}}\left[\sum_{n_1\in\mathbb{Z}} e^{2\pi i n_1 (y^1-\eta_1)} dy^1 \right]\wedge \left[\sum_{n_2\in\mathbb{Z}} e^{2\pi i n_2 (x^2-\eta_2)} dx^2 \right]\wedge \left[\sum_{n_3\in\mathbb{Z}} e^{2\pi i n_3 (x^3-\eta_3)} dx^3 \right]\, ,\\
      \delta(\Pi_{\mathcal{R}\theta\omega})=&\ell_s^3\sum_{\vec{\eta}}\left[\sum_{n_1\in\mathbb{Z}} e^{2\pi i n_1 (x^1-\eta_1)} dx^1 \right]\wedge \left[\sum_{n_2\in\mathbb{Z}} e^{2\pi i n_2 (y^2-\eta_2)} dy^2 \right]\wedge \left[\sum_{n_3\in\mathbb{Z}} e^{2\pi i n_3 (x^3-\eta_3)} dx^3 \right]\, ,
\end{align}
\label{eq: deltas z2z2}
\end{subequations}
where $\vec{\eta}=(\eta_1,\eta_2,\eta_3)$ has entries that are 0 or $\frac{1}{2}$. With all this information, we can then build the three-forms $K_{\a,\eta}$ satisfying \eqref{Ktorus}:
\begin{subequations}
\begin{align}
    K_{\mathcal{R},\eta}=-\ell_s^3\sum_{0\neq \vec{n}\in \mathbb{Z}^3}\frac{e^{2\pi i \vec{n}[(y^1,y^2,y^3)-\vec{\eta}]}}{|\vec{n}|^2} dy^1\wedge dy^2\wedge dy^3\, ,\\
    K_{\mathcal{R}\theta,\eta}=\ell_s^3\sum_{0\neq \vec{n}\in \mathbb{Z}^3}\frac{e^{2\pi i \vec{n}[(x^1,x^2,y^3)-\vec{\eta}]}}{|\vec{n}|^2} dx^1\wedge dx^2\wedge dy^3\, ,\\
    K_{\mathcal{R}\omega,\eta}=\ell_s^3\sum_{0\neq \vec{n}\in \mathbb{Z}^3}\frac{e^{2\pi i \vec{n}[(y^1,x^2,x^3)-\vec{\eta}]}}{|\vec{n}|^2} dy^1\wedge dx^2\wedge dx^3\, ,\\
    K_{\mathcal{R}\theta\omega,\eta}=\ell_s^3\sum_{0\neq \vec{n}\in \mathbb{Z}^3}\frac{e^{2\pi i \vec{n}[(x^1,y^2,x^3)-\vec{\eta}]}}{|\vec{n}|^2} dx^1\wedge dy^2\wedge dx^3\, ,
\end{align}
\end{subequations}
with the indices $\alpha,\eta$  associated to the orientifold planes $\Pi_{\alpha,\eta}$ and $|\vec{n}|^2 = n_1^2/\hat{R}_1^2+n_2^2/\hat{R}_2^2+n_3^2/\hat{R}_3^2$. The relative signs between the different $K_{\alpha}$ are chosen so that $\Im\Omega$ calibrates all the orientifold planes. 

At this stage, we can present the relation in cohomology between the flux $H$ and the orientifold planes derived from \eqref{Krhsum}, so that by using the equations of motion \eqref{intflux} we can fix the complex structure moduli $u_i$. This implies
\begin{equation}
     [\ell_s^{-2}H]=8h \left([\beta^0]-[\beta^1]-[\beta^2]-[\beta^3]\right)\, ,
\end{equation}
where the $\beta^i$ are elements of the following basis of bulk 3-forms:
\bea\nonumber
\a_0 = dx^1 \wedge dx^2 \wedge dx^3\, , & \quad & \b^0 = dy^1 \wedge dy^2 \wedge dy^3 \, ,\\ \nonumber
\a_1 = dx^1 \wedge dy^2 \wedge dy^3\, , & \quad & \b^1 = dy^1 \wedge dx^2 \wedge dx^3 \, ,\\ \nonumber
\a_2 = dy^1 \wedge dx^2 \wedge dy^3\, , & \quad & \b^2 = dx^1 \wedge dy^2 \wedge dx^3 \, ,\\ \nonumber
\a_3 = dy^1 \wedge dy^2 \wedge dx^3\, , & \quad & \b^3 = dx^1 \wedge dx^2 \wedge dy^3 \, .
\eea
Defining $\rho=8\pi^3 \hat{R}_1\hat{R}_2\hat{R}_3$ and considering our choice of complex structure, the holomorphic (3,0)-form $\Omega$ is given by
\bea
\re \Om_{\rm CY} & = &  \ell_s^3 \rho \left(u_1u_2u_3 \b^0 - u_1 \b^1 - u_2 \b^2 - u_3 \b^3 \right) \, ,\\
\im \Om_{\rm CY} & = &  \ell_s^3 \rho \left( \a_0 - u_2u_3 \a_1 - u_1u_3 \a_2 - u_1u_2 \a_3 \right) \, .
\eea
Then, a solution to the first equation in \eqref{intflux} can be accomplished if all the complex structure moduli are fixed to $u_i=1$, and   $\mu=\ell_s^{-1}4h/\rho$.

In light of all this, keeping the complex structure unfixed, we can construct $\mathcal{F}_{\alpha,\eta}=\ell_s d^{\dagger}K_{\alpha,\eta}$. We arrive at:
\begin{subequations}
\begin{align}
    \mathcal{F}_{\mathcal{R},\eta}=\frac{i\ell_s^2}{2\pi}\sum_{0\neq \vec{n}\in \mathbb{Z}^3}\frac{e^{2\pi i \vec{n}[(y^1,y^2,y^3)-\vec{\eta}]}}{|\vec{n}|^2} \left(\frac{n_1}{\hat{R}_1^2}dy^2\wedge dy^3-\frac{n_2}{\hat{R}_2^2}dy^1\wedge dy^3+\frac{n_ 3}{\hat{R}_3^2}dy^1\wedge dy^2\right)\, ,\label{eq: F0 z2z2} \\
    \mathcal{F}_{\mathcal{R}\theta,\eta}=-\frac{i\ell_s^2}{2\pi}\sum_{0\neq \vec{n}\in \mathbb{Z}^3}\frac{e^{2\pi i \vec{n}[(x^1,x^2,y^3)-\vec{\eta}]}}{|\vec{n}|^2} \left(\frac{n_1}{\hat{R}_1^2}dx^2\wedge dy^3-\frac{n_2}{\hat{R}_2^2}dx^1\wedge dy^3+\frac{n_ 3}{\hat{R}_3^2}dx^1\wedge dx^2\right)\, ,\label{eq: F1 z2z2}\\
     \mathcal{F}_{\mathcal{R}\omega,\eta}=-\frac{i\ell_s^2}{2\pi}\sum_{0\neq \vec{n}\in \mathbb{Z}^3}\frac{e^{2\pi i \vec{n}[(y^1,x^2,x^3)-\vec{\eta}]}}{|\vec{n}|^2} \left(\frac{n_1}{\hat{R}_1^2}dx^2\wedge dx^3-\frac{n_2}{\hat{R}_2^2}dy^1\wedge dx^3+\frac{n_ 3}{\hat{R}_3^2}dy^1\wedge dx^2\right)\, ,\\
      \mathcal{F}_{\mathcal{R}\theta\omega,\eta}=-\frac{i\ell_s^2}{2\pi}\sum_{0\neq \vec{n}\in \mathbb{Z}^3}\frac{e^{2\pi i \vec{n}[(x^1,y^2,x^3)-\vec{\eta}]}}{|\vec{n}|^2} \left(\frac{n_1}{\hat{R}_1^2}dy^2\wedge dx^3-\frac{n_2}{\hat{R}_2^2}dx^1\wedge dx^3+\frac{n_ 3}{\hat{R}_3^2}dx^1\wedge dy^2\right)\, .
\end{align}
\label{eq: F z2z2}
\end{subequations}

Finally, we would like to compute $\int \mathcal{F}_{\alpha,\eta}\wedge\mathcal{F}_{\beta,\xi}\wedge J_{CY}$. To perform this integral we regularise it by interchanging the order between summation and integration. The physical interpretation of this procedure corresponds to smearing the O6-plane over a region of radius $\sim \ell_s$, which is the region of $X_6$ where the supergravity approximation cannot be trusted. 
In practice this corresponds to a truncation of the summation over the Fourier modes labelled by $\vec{n}$. In a finite sum we are able to swap summation and integration freely. We then take the limit when the cut-off of the sum diverges, returning to our original system with a localised source. 

At this point we can check some of the statements made in the last section. First of all, we verify that $\Delta_{\alpha,\eta;\alpha,\zeta}=0$. We focus on the simplest case and consider the contribution from two components of  $\Pi^{\rm O6}_\mathcal{R}$. In particular we choose $\alpha=0$ and $\eta=\zeta=(0,0,0)$ and compute
\begin{equation}
    \Delta_{\mathcal{R},\vec{0};\mathcal{R},\vec{0}} = - e^{K/2} \frac{1}{\ell_s^{6}} \int_{\rm X_6}  J_{\rm CY} \wedge \cF_{\mathcal{R},\vec{0}} \wedge \cF_{\mathcal{R},\vec{0}} =0\, .
\end{equation}
Using \eqref{eq: kahler form z2z2} and \eqref{eq: F0 z2z2}  we immediately see that the contribution vanishes, since there is always a wedge product of repeated one-forms. Note that this is independent on the value of $\vec{\eta}$ in \eqref{eq: F0 z2z2}. Therefore we conclude that $\Delta_{\mathcal{R},\eta;\mathcal{R},\zeta}=0$ for any $\eta$ and $\zeta$. Similar cancellations occur for all contributions of this nature involving other cohomology classes. 

We now focus on the remaining possible contributions, which belong to the $\mathcal{N}=2$ sectors of the compactification and are characterised by D6-branes that have similar wrapping numbers in one of the two-tori and different in the other two. For concreteness we consider two examples: one in which the D6-branes intersect over a one-cycle, and one in which there is no intersection. Starting with the former we build the configuration from \eqref{eq: intersecting D6branes example} and evaluate the contribution from the pair of D6-branes associated to  $\Pi^{\rm O6}_\mathcal{R}$ and $\Pi^{\rm O6}_{\mathcal{R}\theta}$. As depicted in figure \ref{fig: z2z2 intersecciones}, the branes intersect over $(T^2)_3$. The associated BIon contribution is 
\begin{align}
     \Delta_{\mathcal{R},\vec{0};\mathcal{R}\theta,\vec{0}}=& - e^{K/2} \frac{1}{\ell_s^{6}} \int_{\rm X_6}  J_{\rm CY} \wedge \cF_{\mathcal{R},\vec{0}} \wedge \cF_{\mathcal{R}\theta,\vec{0}}\nonumber\\
     =&- e^{K/2}\frac{t^3}{4\pi^2\ell_s^6}\int_{X_6} \sum_{0\neq \vec{n},\vec{m}\in\mathbb{Z}^3} \frac{e^{2\pi i\vec{n}(y^1,y^2,y^3)}e^{2\pi i\vec{m}(x^1,x^2,y^3)}}{|\vec{n}|^2|\vec{m}|^2}\frac{n_3m_3}{\hat{R}_3^2}\, \Phi_6\nonumber\\
     =&-e^{K/2}\frac{t^3}{4\pi^2 N_\Gamma}\sum_{0\neq \vec{n},\vec{m}\in\mathbb{Z}^3} \delta_{n_1}\delta_{n_2}\delta_{m_1}\delta_{m_2}\delta_{n_3+m_3}\frac{1}{|\vec{n}|^2|\vec{m}|^2}\frac{n_3m_3}{\hat{R}_3^4}\nonumber\\
     =&\, e^{K/2}\frac{t^3}{4\pi^2 N_\Gamma}\sum_{n_3\neq 0}\frac{n_3^2}{n_3^4}
     = e^{K/2}\frac{t^3}{4\pi^2 N_\Gamma}2\frac{\pi^2}{6} = \frac{T^3_{\rm D4}}{12N_\Gamma}\, ,
     \label{eq: delta z2 intersection}
\end{align}
where we have defined $\Phi_6=\ell_s^6 dx^1\wedge dx^2\wedge dx^3\wedge dy^1\wedge dy^2\wedge dy^3$. To go from the second to the third line we have used the regularisation procedure stated above. It is easy to repeat the same computation for any $ \Delta_{\mathcal{R},\vec\eta;\mathcal{R}\theta,\vec\zeta}$ such that $\eta_3=\zeta_3$ (in order to preserve the intersection along $(T^2)_3$). The new exponential factors arising from \eqref{eq: F z2z2} vanish once the Kronecker deltas are considered. Similarly, the same result is obtained for intersections involving other cohomology classes. Hence, we verify that  an $\mathcal{N}=2$ sector in which  D6-branes intersect over a one-cycle in $(T^2)_i$ contribute as $\frac{T^i}{12N_\Gamma}$ to \eqref{Deltasum}.

Finally we test the case in which the D6-branes do not overlap but run parallel over the one two-torus. To do so, we build the configuration described in \eqref{eq: non intersecting D6branes example} (see figure \ref{fig: z2z2 no intersecciones}) and evaluate the contribution from the D6-brane associated to the $\Pi^{\rm O6}_\mathcal{R}$ and $\Pi^{\rm O6}_{\mathcal{R}\theta}$ as before. We obtain
\begin{align}
     \Delta_{\mathcal{R},\vec{0};\mathcal{R}\theta,(0,0,1/2)}=& - e^{K/2} \frac{1}{\ell_s^{6}} \int_{\rm X_6}  J_{\rm CY} \wedge \cF_{\mathcal{R},\vec{0}} \wedge \cF_{\mathcal{R}\theta,(0,0,1/2)}\nonumber\\
     =&- e^{K/2}\frac{t^3}{4\pi^2 \ell_s^6}\int_{X_6} \sum_{0\neq \vec{n},\vec{m}\in\mathbb{Z}^3} \frac{e^{2\pi i\vec{n}(y^1,y^2,y^3)}e^{2\pi i\vec{m}(x^1,x^2,y^3)}e^{i\pi m_3}}{|\vec{n}|^2|\vec{m}|^2}\frac{n_3m_3}{\hat{R}_3^2} \Phi_6\nonumber\\
     =&-e^{K/2}\frac{t^3}{4\pi^2 N_\Gamma}\sum_{0\neq \vec{n},\vec{m}\in\mathbb{Z}^3} \delta_{n_1}\delta_{n_2}\delta_{m_1}\delta_{m_2}\delta_{n_3+m_3}\frac{(-1)^{m_3}}{|\vec{n}|^2|\vec{m}|^2}\frac{n_3m_3}{\hat{R}_3^4}\nonumber\\
     =&\, e^{K/2}\frac{t^3}{4\pi^2 N_\Gamma}\sum_{n_3\neq 0}\frac{(-1)^{n_3}}{n_3^2}
     = e^{K/2}\frac{t^3}{4\pi^2 N_\Gamma}2\frac{-\pi^2}{12} 
     = -\frac{T^3_{\rm D4}}{24N_\Gamma}\, ,
\end{align}
and so we recover \eqref{interpar}.

It is worth noting that even though \eqref{intercont} and \eqref{interpar} are correct for all the examples we consider, they do not describe the most general scenario we can think of, see footnote \ref{OSL}. For a generic $\mathcal{N}=2$ configuration in which the D6-branes run parallel along the $(T^2)_a$ over one-cycles of length $L$ and separated by a distance $\frac{\ell_s^2t^a}{L} \eta$, one can generalise the computations above to see that the contribution to \eqref{Deltasum} is given in terms of the dilogarithmic function as
\begin{equation}
     \frac{T^a_{\rm D4}}{2\pi^2N_\Gamma}\Re\left[\textrm{Li}_2(e^{2\pi i \eta})\right] = \frac{T^a_{\rm D4}}{N_\Gamma} \oh \left(\frac{1}{6} - \eta(1-\eta) \right)\, .
\end{equation}

\subsection{$T^6/\mathbb{Z}_4$}

Let us now consider the $\mathbb{Z}_4$ orbifold over a factorisable six-torus $T^6=(T^2)_1\times (T^2)_2 \times (T^2)_3$, as discussed in \cite{blumenhagen2000supersymmetric ,blumenhagen2003supersymmetric}, see also \cite{Ihl:2006pp}. The two-dimensional lattice that defines each 2-torus is generated by the basis of the complex plane $e_{i1}=2\pi R_i$ and $e_{i2}=2\pi R_i\tau_i$, where  $R_i$ are the radii of $(T^2)_i$ and $\tau_i=a_i+iu_i$ is its complex structure. The complex coordinate for each 2-torus is
\begin{equation}
    z^i=2\pi R_i(x^i+ \tau_i y^i)\,, \qquad x^i,y^i\in \mathbb{R}\,.
    \label{coordinates}
\end{equation}
The action of the $\mathbb{Z}_4$ group over $T^6$ is generated by an element $\theta$ that acts as follows
\begin{equation}
    \theta(z^i)=e^{2\pi i v_i} z^i\,,
\end{equation}
with $v_i=(1/4,1/4,1/2)$. The action of this group severely constrains the complex structure. In fact, the complex structure of the first two $T^2$'s is fixed. For the third torus, in which the $\mathbb{Z}_4$ action has an orbit of order $2$, the constrains are less severe. There are two options available, commonly denoted by  AAA and AAB \cite{blumenhagen2003supersymmetric,forste2001orientifolds}, and both of them have $u_3$ free. The AAA case is characterised by the choice $a_3=0$, whereas the $AAB$ has $a_3=1/2$. Therefore, in the $\mathbb{Z}_4$ orbifold  there is always one unconstrained complex structure modulus.

For concreteness let us consider the choice AAA. All the steps of the analysis can be replicated in the AAB scenario to arrive to the same results. In the present case, we have $\tau_1=\tau_2=i$ and $\tau_3=i u_3$. The basis of the lattice that generates the torus is orthogonal and gives the following identifications 
\begin{subequations}
\begin{align}
    z^1 &\sim z^1+2\pi R_1 \sim z^1+2\pi i R_1\, ,\\
    z^2 &\sim z^2+2\pi R_2 \sim z^2+2\pi i R_2\, ,\\
    z^3 &\sim z^3+2\pi R_3 \sim z^3+2\pi i u_3 R_3\, .
\end{align}
\label{eq: periodic identification Z4}
\end{subequations}
Up to the constraints on the complex structure, the covering space metric and the Kähler form are the same as in the $\mathbb{Z}_2\times \mathbb{Z}_2$ case.
\begin{align}
    g =&\, 4\pi^2 \ell_s^2\, {\rm diag} \left(\hat{R}_1^2, \hat{R}_2^2, \hat{R}_3^2, \hat{R}_1^2, \hat{R}_2^2, u_3^2\hat{R}_3^2 \right)\,   ,  \\
    J=&\, \ell_s^2( t^1 dx^1\wedge dy^1 +t^2 dx^2\wedge dy^2 +t^3 dx^3\wedge dy^3)\, ,
    \label{eq: J z4}
\end{align}
where again we  defined the dimensionless radii $\hat{R}_i=R_i/\ell_s$ and the K\"ahler moduli $t^i=4\pi^2 \hat{R}_{i}^2 u_i$.

\begin{figure}[h!]
  \includegraphics[width=\textwidth]{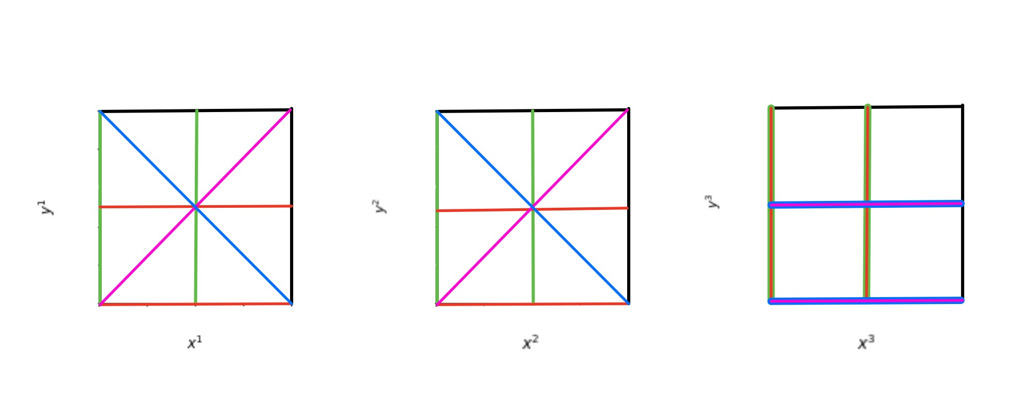}
    \caption{Orientifold planes projected over $T^2\times T^2\times T^2$ in the $\mathbb{Z}_4$ orbifold. Planes $\Pi_\mathcal{R}$, $\Pi_{1}$, $\Pi_{2}$, $\Pi_{3}$ are represented by the colours red, pink, green and blue respectively.}
    \label{figure: Z4-planes}
\end{figure}

The orientifold planes are given by the fixed points of the orientifold involution $\sigma(z)=\bar{z}$, up to orbifold action identifications. Consequently, we have the following orientifold planes, summarised in table \ref{table: z4 orientifolds} and represented in figure \ref{figure: Z4-planes}.
\renewcommand{\arraystretch}{0.9}
\begin{table}[H]
$$
\begin{array}{|l|l|c|}
\hline \Pi_{\alpha} & \text { Fixed point equation } & \text { O6-plane position } \\
\hline \Pi_{0} & \sigma\left(z^{a}\right)=z^{a} & y^{1} \in\left\{0,\frac{1}{2}\right\} \quad y^{2} \in\left\{0,\frac{1}{2}\right\} \quad y^{3} \in\left\{0,\frac{1}{2}\right\} \\
\Pi_{1} & \sigma\left(z^{a}\right)=\theta\left(z^{a}\right) & y^{1}- x^{1} =0 \quad y^{2}- x^{2} =0 \quad x^{3} \in\{0,\oh\} \\
\Pi_{2} & \sigma\left(z^{a}\right)=\theta^{2}\left(z^{a}\right) & x^1\in\{0,\oh\} \quad x^{2}\in\{0,\oh\} \quad y^{3}\in\{0,\oh\} \\
\Pi_{3} & \sigma\left(z^{a}\right)= \theta^{3}\left(z^{a}\right) & y^{1} +x^1=1 \quad y^{2}+ x^{2} =1 \quad x^{3}\in\{0,\oh\} \\
\hline
\end{array}
$$
\caption{O6-planes in $T^6/\mathbb{Z}_4$.}
\label{table: z4 orientifolds}
\end{table}
The above content of O6-planes can be expressed in terms of invariant bulk three-cycles following the same reasoning as in the $\IZ_2\times\IZ_2$ case. Let $\pi_{2i-1}$ and $\pi_{2i}$ constitute a basis of fundamental one-cycles on the torus $T^2_i$ $(i=1,2,3)$, i. e. cycles winded once along the periodic directions given by the identifications that defined our tori in \eqref{eq: periodic identification Z4}. We  used them to build the following three-cycles
\be
\pi_{IJK}=\pi_I \otimes \pi_J \otimes \pi_K\, ,
\ee
with $I=1,2$, $J=3,4$ and $K=5,6$. For the $\IZ_4$ orientifold the minimal invariant bulk three-cycles are given by \cite{blumenhagen2003supersymmetric}
\begin{equation}
\begin{array}{ll}
\rho_{1} \equiv 2\left(\pi_{135}-\pi_{245}\right), & \bar{\rho}_{1} \equiv 2\left(\pi_{136}-\pi_{246}\right)\, , \\
\rho_{2} \equiv 2\left(\pi_{145}+\pi_{235}\right), & \bar{\rho}_{2} \equiv 2\left(\pi_{146}+\pi_{236}\right).
\end{array} \label{eq: bulk cycles z4}
\end{equation}
The factor of 2 in \eqref{eq: bulk cycles z4} is due to the fact that $\theta^2$ acts trivially over $\pi_{ijk}$. Hence, the O6-planes content can be expressed as
\be
\Pi_{\rm O6}=4\rho_1 - 2\bar{\rho}_2\, .
\ee

As we have seen, due to the factorised structure of the orbifold, the orientifold three-cycles are also factorised as products of one-cycles in the covering space, each one defined in each of the two-tori. A $\delta$-function supported on these one-dimensional objects can be expressed using the conventional Fourier expansion for the $\delta$-function distribution:
\begin{equation}
    \delta(w)=\frac{1}{S}\sum_{n\in\mathbb{Z}}e^{2\pi i n w/S}\, ,
    \label{eq: fourier delta def}
\end{equation}
where $w$ denotes the direction transverse to the cycle normalised to unit norm and $S$ is the periodicity of the configuration along such a transverse direction. Therefore, in order to build the bump $\d$-functions for factorisable three-cycles, we need to find the transverse periodicity $S$ of the respective one-cycles, which we define as  the distance that separates two consecutive intersection points between the loci of the cycle (given by the linear equations of table \ref{table: z4 orientifolds}) projected over the two-torus we are considering  and the transverse direction to the cycle in that same two-torus. As a general rule, if we have a minimal-length one-cycle of length $L$ on a two-torus of area $A$, the dimensionless transverse period $S$ that appears in \eqref{eq: fourier delta def} will be $S=A /\ell_s L$.

We did not have to worry about this factor in the $\mathbb{Z}_2\times \mathbb{Z}_2$ example, since all the cycles had periodicity one in the normalised coordinates. That will no longer be the case in general for the rest of our examples. We illustrate this reasoning by building the $\delta$-like bump functions with support on to the loci $\Pi_i$ introduced in table \ref{table: z4 orientifolds}. The factor $S$ will be crucial to properly define the $\delta$-bump function describing the orientifold planes that do not decompose as a single product of fundamental one-cycles, such as $\Pi_1$.
\begin{subequations}
\begin{align}
    \delta(\Pi_{0})=&\ell_s^3\sum_{\vec{\eta}}\left[\sum_{n_1\in\mathbb{Z}} e^{2\pi i n_1 (y^1-\eta_1)} dy^1 \right]\wedge \left[\sum_{n_2\in\mathbb{Z}} e^{2\pi i n_2 (y^2-\eta_2)} dy^2 \right]\wedge \left[\sum_{n_3\in\mathbb{Z}} e^{2\pi i n_3 (y^2-\eta_2)} dy^3 \right]\, ,\\
    \delta(\Pi_{1})=&\ell_s^3\left[\sqrt{2}\sum_{n_1\in\mathbb{Z}} e^{2\sqrt{2}\pi i n_1 \hat{y}^1} d\hat{y}^1 \right]\wedge \left[\sqrt{2}\sum_{n_2\in\mathbb{Z}} e^{2\sqrt{2}\pi i n_2 \hat{y}^2} d\hat{y}^2 \right]\wedge \left[\sum_{\eta_3}\sum_{n_3\in\mathbb{Z}} e^{2\pi i n_3 (x^3-\eta_3)} dx^3 \right]\, ,\\
    \delta(\Pi_{2})=&\ell_s^3\sum_{\vec{\eta}}\left[\sum_{n_1\in\mathbb{Z}} e^{2\pi i n_1 (x^1-\eta_1)} dx^1 \right]\wedge \left[\sum_{n_2\in\mathbb{Z}} e^{2\pi i n_2 (x^2-\eta_2)} dx^2 \right]\wedge \left[\sum_{n_3\in\mathbb{Z}} e^{2\pi i n_3 (y^3-\eta_3)} dy^3 \right]\, ,\\
    \delta(\Pi_{3})=&\ell_s^3\left[\sqrt{2}\sum_{n_1\in\mathbb{Z}} e^{2\sqrt{2}\pi i n_1 \tilde{y}^1} d\tilde{y}^1 \right]\wedge \left[\sqrt{2}\sum_{n_2\in\mathbb{Z}} e^{2\sqrt{2}\pi i n_2 \tilde{y}^2} d\tilde{y}^2 \right]\wedge \left[\sum_{\eta_3}\sum_{n_3\in\mathbb{Z}} e^{2\pi i n_3 (x^3-\eta_3)} dx^3 \right]\, ,
\end{align}
\end{subequations}
where we have defined $\hat{y}^i=\frac{1}{\sqrt{2}}(x^i-y^i)$, $\tilde{y}^i=\frac{1}{\sqrt{2}}(x^i+y^i)$ and $\vec{\eta}$ has entries that are either $0$ or $1$. With all this information it is straightforward to build the three-form $K$ satisfying \eqref{eq: K equation} through the introduction of the following set of three-form currents defined in \eqref{Ktorus}:

\begin{subequations}
\begin{align}
    &K_{0,\eta}=-\ell_s^3\sum_{0\neq \vec{n}\in \mathbb{Z}^3}\frac{e^{2\pi i \vec{n}[(y^1,y^2,y^3)-\vec{\eta}]}}{|\vec{n}|^2} dy^1\wedge dy^2\wedge dy^3\, ,\\
    &K_{1,\eta_3}=-2\ell_s^3\sum_{0\neq \vec{n}\in \mathbb{Z}^3}\frac{e^{2\pi i \vec{n}[(\sqrt{2}\hat{y}^1,\sqrt{2}\hat{y}^2,x^3)-(0,0,\eta_3)]}}{|\vec{n}|^2} d\hat{y}^1\wedge d\hat{y}^2\wedge dx^3\,,\\
    &K_{2,\eta}=\ell_s^3\sum_{0\neq \vec{n}\in \mathbb{Z}^3}\frac{e^{2\pi i \vec{n}[(x^1,x^2,y^3)-\vec{\eta}]}}{|\vec{n}|^2} dx^1\wedge dx^2\wedge dy^3\, ,\\
   &K_{3,\eta_3}=2\ell_s^3\sum_{0\neq \vec{n}\in \mathbb{Z}^3}\frac{e^{2\pi i \vec{n}[(\sqrt{2}\tilde{y}^1,\sqrt{2}\tilde{y}^2,x^3)-(0,0,\eta_3)]}}{|\vec{n}|^2} d\tilde{y}^1\wedge d\tilde{y}^2\wedge dx^3\, ,
\end{align}
\end{subequations}
 where $K_{\alpha,\eta}$ is the function associated to $\Pi_{\alpha ,\eta}$ and  $|\vec{n}|^2=n_1^2/S_1^2 \hat{R}_1^2+n_2^2/S_2^2\hat{R}_2^2+n_3^2/S_3^2\hat{R}_3^2$, with $S_i$ the transverse period of the one-cycle obtained from projecting the three-cycle $\Pi$ over $(T^2)_i$. Note that $|\vec{n}|$ changes for each function $K_{\alpha}$, since each one is describing a different three-cycle. Also, as before, the relative signs between the different $K_{\alpha}$ are chosen so that $\Im\Omega$ calibrates all the orientifold planes.

At this point, we introduce the cohomology relation $[\ell_s^{-2}H]=h\rm P.D[\Pi_{\rm O6}]$, which implies
\begin{equation}
\begin{split}
     \ell_s^{-2} [H]=&h\left( 8[\beta^0]  +4[d\hat{y}^{1}\wedge d\hat{y}^{2}\wedge dx^3] - 4[d\tilde{y}^{1}\wedge d\tilde{y}^{2}\wedge dx^3]  - 8[\beta^3]\right) \\
      = &\ 8h\left([\beta^0]-\frac{1}{2}[\beta^1]-\frac{1}{2}[\beta^2]-[\beta^3]\right)\, .        
\end{split}
\end{equation}
Now we can impose the equation of motion using \eqref{intflux}. Defining $\rho=8\pi^3 \hat{R}_1\hat{R}_2\hat{R}_3$ and taking into account our choice of complex structure, the holomorphic form $\Omega$ is
\bea
\re \Om_{\rm CY} & = &  \ell_s^3 \rho \left(u_3 \b^0 -  \b^1 - \b^2 - u_3 \b^3 \right) \, ,\\
\im \Om_{\rm CY} & = &  \ell_s^3 \rho \left( \a_0 -  u_3\a_1 - u_3 \a_2 -  \a_3 \right) \, .
\eea
In order to satisfy \eqref{intflux}  the remaining complex structure modulus must be fixed to $u_3=2$, while $\mu$ is given by $\mu=\ell_s^{-1}4h/\rho u_3$. 

Along the lines of the $\IZ_2\times\IZ_2$ case, let us turn to the appropriate flux quantisation condition in the $\IZ_4$ orientifold. Taking the results from \cite{blumenhagen2003supersymmetric}, the minimal integral lattice of three-cycles is defined as in \eqref{eq: bulk cycles z4}. Applying the flux quantisation criterium for the $H$ flux once we consider the presence of O6-planes we find that $[\ell_s^{-2}H] =2h{\rm P.D} [2\rho_1 - \bar{\rho}_2] = h{\rm P.D} [\Pi_{\rm O6}]$ with $h\in \IZ$.

This quantisation condition is more constraining than in the $\IZ_2 \times \IZ_2$ orbifold, allowing solutions to the tadpole involving only a single jump in the quantum of Roman mass 
\begin{equation}
    m = 2k\, , \qquad h = 1\, , \qquad N=4-2k\,,\qquad k = 1,2. \label{eq: tadpole z4}
\end{equation}

Next, we compute the different components of $\mathcal{F}$ in \eqref{eq: worldvolume flux} as $\mathcal{F}_{\alpha,\eta}=\ell_s d^\dagger K_{\alpha,\eta}$:
\begin{subequations}
\begin{align}
   \mathcal{F}_{0,\eta}=&\frac{i\ell_s^2}{2\pi}\sum_{0\neq \vec{n}\in \mathbb{Z}^3}\frac{e^{2\pi i \vec{n}[(y^1,y^2,y^3)-\vec{\eta}]}}{|\vec{n}|^2} \left(\frac{n_1}{\hat{R}_1^2}dy^2\wedge dy^3-\frac{n_2}{\hat{R}_2^2}dy^1\wedge dy^3+\frac{n_ 3}{\hat{R}_3^2}dy^1\wedge dy^2\right)\, , \label{eq: F0 Z4}\\
   \mathcal{F}_{1,\eta}=&\frac{i\ell_s^2}{2\pi}2\sum_{0\neq \vec{n}\in \mathbb{Z}^3}\frac{e^{2\pi i \vec{n}[(\sqrt{2}\hat{y}^1,\sqrt{2}\hat{y}^2,x^3)-(0,0,\eta_3)]}}{|\vec{n}|^2} \left(\frac{\sqrt{2}n_1}{\hat{R}_1^2}d\hat{y}^2\wedge dx^3-\frac{\sqrt{2}n_2}{\hat{R}_2^2}d\hat{y}^1\wedge dx^3+\frac{n_ 3}{\hat{R}_3^2}d\hat{y}^1\wedge d\hat{y}^2\right)\, , \\
    \mathcal{F}_{2,\eta}=&-\frac{i\ell_s^2}{2\pi}\sum_{0\neq \vec{n}\in \mathbb{Z}^3}\frac{e^{2\pi i \vec{n}[(x^1,x^2,y^3)-\vec{\eta}]}}{|\vec{n}|^2} \left(\frac{n_1}{\hat{R}_1^2}dx^2\wedge dy^3-\frac{n_2}{\hat{R}_2^2}dx^1\wedge dy^3+\frac{n_ 3}{\hat{R}_3^2}dx^1\wedge dx^2\right)\, , \\
    \mathcal{F}_{3,\eta}=&-\frac{i\ell_s^2}{2\pi}2\sum_{0\neq \vec{n}\in \mathbb{Z}^3}\frac{e^{2\pi i \vec{n}[(\sqrt{2}\tilde{y}^1,\sqrt{2}\tilde{y}^2,x^3)-(0,0,\eta_3)]}}{|\vec{n}|^2} \left(\frac{\sqrt{2}n_1}{\hat{R}_1^2}d\tilde{y}^2\wedge dx^3-\frac{\sqrt{2}n_2}{\hat{R}_2^2}d\tilde{y}^1\wedge dx^3+\frac{n_ 3}{\hat{R}_3^2}d\tilde{y}^1\wedge d\tilde{y}^2\right)\, .\label{eq: F3 Z4}
\end{align}
\end{subequations}
Now we would like to compute $\int J_\cy\wedge \mathcal{F}_{\alpha,\eta}\wedge\mathcal{F}_{\beta,\zeta}$. To perform this integral we regularise it as before, interchanging the order between summation and integration.  Similarly to the $\mathbb{Z}_2\times \mathbb{Z}_2$ orbifold, this allows us to obtain Kronecker deltas from the following relations:
\begin{align}
    \int_{T^2}  e^{2\pi i ny^1}e^{2\pi^i my^1} dx^1 dy^1=&\, \delta_{n+m}\, ,\\
    \int_{T^2} e^{2\sqrt{2}\pi i n\tilde{y}^1}e^{2\sqrt{2}\pi^i m\tilde{y}^1} dx^1 dy^1=&\, \delta_{n+m}\, ,\\
    \int_{T^2}  e^{2\sqrt{2}\pi i n\hat{y}^1}e^{2\sqrt{2}\pi^i m\hat{y}^1} dx^1 dy^1=&\, \delta_{n+m}\, ,\\
    \int_{T^2} e^{2\pi i ny^1}e^{2\sqrt{2}\pi^i m\tilde{y}^1} dx^1 dy^1=&\, \delta_n\delta_m\, ,\\
    \int_{T^2} e^{2\pi i ny^1}e^{2\sqrt{2}\pi^i m\hat{y}^1} dx^1 dy^1=&\, \delta_n\delta_m\, ,\\
    \int_{T^2} e^{2\sqrt{2}\pi i n\tilde{y}^1}e^{2\sqrt{2}\pi^i m\hat{y}^1} dx^1 dy^1=&\, \delta_n\delta_m\, .
\end{align}
With all this information we can finally evaluate the different terms that contribute to \eqref{Deltasum}. Many of them will be exactly as in the $\mathbb{Z}_2\times \mathbb{Z}_2$ orbifold, but there are also some new kinds of  contributions. First of all, we can consider pairs of three-cycles with non-vanishing intersection number. Let us for instance choose  $\Delta_{0,\vec{0};3,\vec{0}}$. From figure \ref{figure: Z4-planes} we see that the three-cycles intersect at a single point. Using \eqref{eq: J z4}, \eqref{eq: F0 Z4} and \eqref{eq: F3 Z4} we obtain

\begin{align}
\Delta_{0,\vec{0};3,0}=& -  \frac{e^{K/2}}{\ell_s^{6}} \int_{\rm X_6}  J_{\rm CY} \wedge \cF_{0,\vec{0}} \wedge \cF_{3,0}\nonumber\\
=&- \frac{e^{K/2}}{4\pi^2\ell_s^{6}}\int_{X_6}\sum_{0\neq \vec{n},\vec{m}\in \mathbb{Z}^3}\frac{2e^{2\pi i [\vec{n}(y^1,y^2,y^3)+ \vec{m}(\sqrt{2}\tilde{y}^1,\sqrt{2}\tilde{y}^2,x^3)]}}{|\vec{n}|^2|\vec{m}|^2}\left(\frac{n_1m_1t^1}{\hat{R}_1^4}+\frac{n_2m_2 t^2}{\hat{R}_2^4}+\frac{n_3m_3t^3}{2\hat{R}_3^4}\right)\Phi_6\nonumber\\
=&-e^{K/2}\frac{1}{4\pi^2N_\Gamma} \sum_{0\neq \vec{n},\vec{m}\in \mathbb{Z}^3}\frac{1}{|\vec{n}|^2|\vec{m}|^2}\left(\frac{2n_1m_1t^1}{\hat{R}_1^4}+\frac{2n_2m_2 t^2}{\hat{R}_2^4}+\frac{n_3m_3t^3}{\hat{R}_3^4}\right)\delta_{n_1}\delta_{n_2}\delta_{n_3}\delta_{m_1}\delta_{m_2}\delta_{m_3}\nonumber\\
=&\, 0   \, ,
\end{align}
where we defined $\Phi_6=\ell_s^6 dx^1\wedge dx^2\wedge dx^3\wedge dy^1\wedge dy^2\wedge dy^3$. Therefore, we observe once more that the only non-trivial contributions come from the $\mathcal{N}=2$ sector. For the case of $\mathbb{Z}_4$ orbifold, the aforementioned sector is richer and more diverse than the $\mathbb{Z}_2\times\mathbb{Z}_2$ orbifold. In addition to pairs of branes of the form \eqref{eq: delta z2 intersection} we must also consider contributions involving cycles that do not run along the fundamental periodic directions. Let us focus on $\Delta_{1,0;3,0}$. In figure \ref{figure: Z4-planes} we can observe the involved three-cycles intersect over a one-cycle on $(T^2)_3$. We find that
\begin{align}
\Delta_{1,0;3,0}=& - e^{K/2} \frac{1}{\ell_s^{6}} \int_{\rm X_6}  J_{\rm CY} \wedge \cF_{1,0} \wedge \cF_{3,0}\nonumber\\
=&-e^{K/2} \frac{1}{4\pi^2\ell_s^{6}}\int_{X_6}\sum_{0\neq \vec{n},\vec{m}\in \mathbb{Z}^3}\frac{e^{2\pi i \vec{n}(\sqrt{2}\hat{y}^1,\sqrt{2}\hat{y}^2,x^3)}}{|\vec{n}|^2}\frac{e^{2\pi i \vec{m}(\sqrt{2}\tilde{y}^1,\sqrt{2}\tilde{y}^2,x^3)}}{|\vec{m}|^2}\frac{4n_3m_3t^3}{\hat{R}_3^4}\Phi_6\nonumber\\
=&-e^{K/2}\frac{1}{4\pi^2N_\Gamma} \sum_{0\neq \vec{n},\vec{m}\in \mathbb{Z}^3}\frac{1}{|\vec{n}|^2|\vec{m}|^2}\frac{4n_3m_3t_3}{\hat{R}_3^4}\delta_{n_1}\delta_{n_2}\delta_{m_1}\delta_{m_2}\delta_{n_3+m_3}\nonumber\\
=&-e^{K/2}\frac{1}{4\pi^2N_\Gamma} \sum_{0\neq n_3}\frac{\hat{R}_3^4}{4n_3^4}\frac{-4n_3^2t_3}{\hat{R}_3^4}=\frac{T^3_{\rm D4}}{12N_\Gamma}\, .
\end{align}
 The result again agrees with \eqref{intercont}. Similarly, we can consider cycles that do not intersect, but run parallel along one of the two-torus. We take, for instance, $\mathcal{F}_{1,0}$ and $\mathcal{F}_{3,1/2}$, obtaining
\begin{align}
     \Delta_{1,0;3,1/2}=& - e^{K/2} \frac{1}{\ell_s^{6}} \int_{\rm X_6}  J_{\rm CY} \wedge \cF_{1,0} \wedge \cF_{3,1/2}\nonumber\\
     =&-e^{K/2} \frac{1}{4\pi^2\ell_s^{6}}\int_{X_6}\sum_{0\neq \vec{n},\vec{m}\in \mathbb{Z}^3}\frac{e^{2\pi i \vec{n}(\sqrt{2}\hat{y}^1,\sqrt{2}\hat{y}^2,x^3)}}{|\vec{n}|^2}\frac{e^{2\pi i \vec{m}(\sqrt{2}\tilde{y}^1,\sqrt{2}\tilde{y}^2,x^3)}e^{i\pi m_3}}{|\vec{m}|^2}\frac{4n_3m_3t^3}{\hat{R}_3^4}\Phi_6\nonumber\\
    =&-e^{K/2}\frac{1}{4\pi^2N_\Gamma} \sum_{0\neq \vec{n},\vec{m}\in \mathbb{Z}^3}\frac{(-1)^{m_3}}{|\vec{n}|^2|\vec{m}|^2}\frac{4n_3m_3t_3}{\hat{R}_3^4}\delta_{n_1}\delta_{n_2}\delta_{m_1}\delta_{m_2}\delta_{n_3+m_3}\nonumber\\
    =&e^{K/2}\frac{1}{4\pi^2N_\Gamma} \sum_{0\neq n_3}\frac{(-1)^{n_3}t^3}{n_3^2}=-\frac{T^3_{\rm D4}}{24N_\Gamma}\, .
\end{align}
Putting all the contributions together we conclude that
\begin{equation}
    \Delta_{\rm D8}^{\rm BIon} = \frac{1}{24N_\Gamma}\, \left(\sum_{\alpha,\beta}\hat{q}_{0,\alpha}\,\hat{q}_{2,\beta}\,\varepsilon_{\alpha\beta}  +\sum_{\sigma,\rho}4\hat{q}_{1,\sigma}\,\hat{q}_{3,\rho}\,\varepsilon_{\sigma\rho}\right)T^3_{\rm{D4}} \, ,
\end{equation}
with $\varepsilon_{\alpha\beta}$ defined as in \eqref{finalDelta}. Taking as an example the family of solutions defined in \eqref{eq: tadpole z4} we can provide again a configuration of D6-branes with negative $\Delta_{\rm D8}^{\rm Bion}$. For instance, let us consider a configuration such that for each value of $\alpha$ all the corresponding $p_{\alpha}(4-2k)$ D6-branes are wrapping a single three-cycle. In particular, one can take
\begin{equation}
    \hat{q}_{0,(0,0,0)}=\hat{q}_{2,(0,0,\frac{1}{2})}= 8, \hspace{3em} \hat{q}_{1,(0,0,0)}=\hat{q}_{3,(0,0,\frac{1}{2})}=2.
\end{equation}
With such a configuration we obtain
\begin{equation}
    \Delta_{\rm D8}^{\rm Bion}= -\frac{5}{6} T^{3}_{\rm D4}\, .
\end{equation}
Therefore, this result signals again an BIonic excess tension for the 4d membrane, which could imply a possible failure of the WGC inequality. Indeed, a naive computation\footnote{For all the $\mathbb{Z}_4$ orbifolds studied in \cite[Appendix B]{Reffert:2006du} one obtains the relations $c_2(X_6).R_i =0$, $R_i \simeq 4D_{i\a} +\dots$ $i=1,2$ and $c_2(X_6).R_3 =0$, $R_3 \simeq 2D_{3\a} +\dots$, where the dots represent exceptional divisors. From here one can deduce that $\Delta_{\rm D8}^{\rm curv} = \oh  T^{3}_{\rm D4}$, following the same reasoning as in the $\Z_2 \times Z_2$ orbifold.} gives 
$\Delta_{\rm D8}^{\rm curv} = \oh  T^{3}_{\rm D4}$ in the orbifold limit, which implies that  $\Delta_{\rm D8}^{\rm curv} + \Delta_{\rm D8}^{\rm Bion} < 0$. Hence, this vacuum also seems to be in tension with the WGC for 4d membranes. 

Repeating the analysis for the choice AAB provides the same results.

\subsection{$T^6/\mathbb{Z}_3\times \mathbb{Z}_3$}

We consider now the case where the internal space is an orientifold of the orbifold $T^6/\mathbb{Z}_3^2$ described in \cite{strominger1985topology,DeWolfe:2005uu,Junghans:2020acz}. In order to be consistent with our choice of orientifold involution, we will slightly change the notation of the aforementioned references.

We will again work in the covering space, which is a factorisable six torus, $T^6=(T^2)_1\times (T^2)_2 \times (T^2)_3$   with complex coordinates $z^i$ given by \eqref{coordinates}. 
The $\mathbb{Z}_3\times\mathbb{Z}_3$ orbifold action reads 
\begin{equation}
    \theta: z^i \rightarrow \alpha^2 z^i\,,\qquad  \omega:z^i\rightarrow \alpha^{2i}z^i+\left(\frac{1}{2}i+\frac{\sqrt{3}}{2}\right)\, ,
\end{equation}
with $\alpha=e^{i\pi/3}$. 

The above symmetries, together with the orientifold involution, are more constraining that those introduced in the $\mathbb{Z}_2\times \mathbb{Z}_2$
or $\mathbb{Z}_4$ orbifolds and they fully fix the complex structure to
\begin{equation}
    \tau_1=\tau_2=\tau_3=\frac{\sqrt{3}}{2}+\frac{1}{2}i\, .
\end{equation}
Hence, the factor $(T^2)_i$ can be described as a quotient of $\mathbb{C}$ by a lattice generated by $e_{i1}=2\pi R_i i$ and $e_{i2}=2\pi R_i(\sqrt{3}/2+i/2)$.  They provide the following periodic identifications:
\begin{equation}
    z^i\sim z^i+2\pi R_i i \sim z^i+2\pi R_i \left(\frac{1}{2}i+\frac{\sqrt{3}}{2}\right)\,.
    \label{eq: periodic identification Z3z3}
\end{equation}

It is worth noting that only the  generator $\theta$ of the first $\mathbb{Z}_3$ preserves the lattice generated by these vectors. The trick of this orbifold is that we are not taking the quotient simultaneously. $Q$ is not a symmetry of $T^6$ by itself, but it emerges as a symmetry of the quotient $T^6/\mathbb{Z}_3^{\theta}$. This construction was described in detail in \cite{strominger1985topology}.  Using the periodic coordinates, the metric and the Kähler form are 
\begin{align}
    g =&\, 4\pi^2\ell_s^2\left(\begin{array}{cccccc}
        \hat{R}_1^2 & 0 & 0 & \frac{\hat{R}_1^2}{2} & 0 & 0 \\
        0 & \hat{R}_2^2 & 0 & 0 &  \frac{\hat{R}_2^2}{2} & 0 \\
        0 & 0 & \hat{R}_3^2 & 0 & 0 &  \frac{\hat{R}_3^2}{2} \\
         \frac{\hat{R}_1^2}{2} & 0 & 0 & \hat{R}_1^2 & 0 & 0 \\
        0 &  \frac{\hat{R}_2^2}{2} & 0 & 0 & \hat{R}_2^2 & 0 \\
        0 & 0 &  \frac{\hat{R}_3^2}{2} & 0 & 0 & \hat{R}_3^2 
    \end{array}\right)\,,\\
    \nonumber\\
    J=&\, \ell_s^2(t^1 dx^1\wedge dy^1 +t^2 dx^2\wedge dy^2 + t^3 dx^3\wedge dy^3)\,,\label{eq: kahler form z3z3}    
\end{align}
where  we  defined the dimensionless radii $\hat{R}_i=R_i/\ell_s$ and the K\"ahler moduli $t^i=4\pi^2\sqrt{3}/2 \hat{R}_{i}^2$.

\begin{figure}[h!]
  \includegraphics[width=\textwidth]{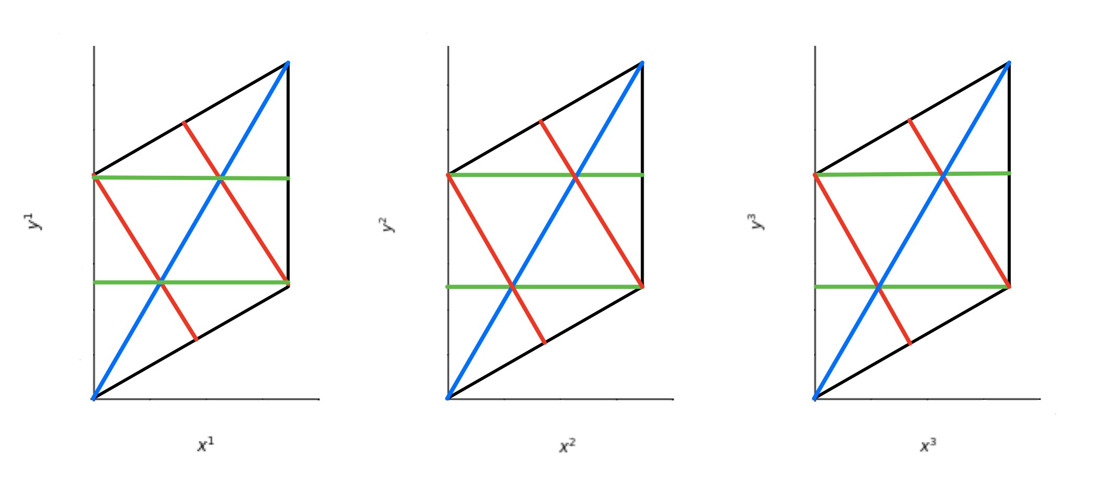}
    \caption{Fundamental domain of $T^2\times T^2\times T^2$ and the fixed loci for $T^6/\mathbb{Z}_3\times \mathbb{Z}_3$. Planes $\Pi_0$, $\Pi_{1}$, $\Pi_{2}$ are represented by the colours green, red and blue respectively. }
    \label{figure: tilted torus}
\end{figure}

The orientifold planes are given by the set of points that satisfy $\sigma(z)=z$ up to the action of the orbifold. This gives nine different loci, summarised in table \ref{table: z3xz3 orientifolds} and represented schematically in figure \ref{figure: tilted torus}. Using the expression $S=A /\ell_s L$, it is easy to see that the transverse period of all the one-cycles involved in the problem is $S=1/2$. 

\renewcommand{\arraystretch}{0.95}
\begin{table}[H]
$$
\begin{array}{|l|l|lll|}
\hline \Pi_{\alpha} & \text { Fixed point equation } & & \text { O6-plane position }  & \\
\hline \Pi_{0} & \sigma\left(z^{a}\right)=z^{a} & x^1+2y^{1} \in\left\{1, 2\right\} \quad &x^2+2y^{2} \in\left\{1, 2\right\} &x^3+2y^{3} \in\left\{1, 2\right\} \\
\Pi_{1} & \sigma\left(z^{a}\right)=\theta\left(z^{a}\right) & y^{1}+2 x^{1} \in\{1,2\} &y^{2}+2 x^{2} \in\{1,2\} &y^{3}+2 x^{3} \in\{1,2\} \\
\Pi_{2} & \sigma\left(z^{a}\right)=\theta^{2}\left(z^{a}\right) & y^{1}- x^{1}=0 &y^{2}- x^{2}=0 &y^{3}- x^{3}=0 \\
\Pi_{3} & \sigma\left(z^{a}\right)=\omega \theta^{2}\left(z^{a}\right) & x^1+2y^{1} \in\left\{1, 2\right\} &y^{2}+2 x^{2} \in\{1,2\} &y^{3}-x^{3}=0 \\
\Pi_{4} & \sigma\left(z^{a}\right)=\omega^{2} \theta\left(z^{a}\right) & x^1+2y^{1} \in\left\{1, 2\right\} &y^{2}- x^{2}=0 &y^{3}+2 x^{3} \in\{1,2\} \\
\Pi_{5} & \sigma\left(z^{a}\right)=\omega \theta\left(z^{a}\right) & y^{1}- x^{1}=0 &x^2+2 y^{2} \in\left\{1, 2\right\} &y^{3}+2 x^{3} \in\{1,2\} \\
\Pi_{6} & \sigma\left(z^{a}\right)=\omega^{2} \theta^{2}\left(z^{a}\right) & y^{1}+2 x^{1} \in\{1,2\} &x^2+2y^{2} \in\left\{1, 2\right\} &y^{3}- x^{3}=0 \\
\Pi_{7} & \sigma\left(z^{a}\right)=\omega\left(z^{a}\right) & y^{1}+2 x^{1} \in\{1,2\} &y^{2}- x^{2}=0 &x^3+2y^{3} \in\left\{1, 2\right\} \\
\Pi_{8} & \sigma\left(z^{a}\right)=\omega^{2}\left(z^{a}\right) & y^{1}- x^{1}=0 & y^{2}+2 x^{2} \in\{1,2\} & x^3+2 y^{3} \in\left\{1, 2\right\} \\
\hline
\end{array}
$$
\caption{O6-planes in $T^6/\mathbb{Z}_3\times \mathbb{Z}_3$.}
\label{table: z3xz3 orientifolds}
\end{table}

The above O6-plane content can be expressed in terms of bulk three-cycles $\rho_i$. Consider again the three-cycles inherited from the covering space $T^6$. Let us define the basis of fundamental one-cycles $\pi_{2i-1}$ and $\pi_{2i}$ of the tilted torus $(T^2)_i$, i.e. cycles winded once along the periodic directions given by the identifications that defined our tori in \eqref{eq: periodic identification Z3z3}.

Then, summing over the orbits of two three-cycles, say $\pi_{135}$ and $\pi_{136}$, we obtain the following two invariant three-cycles $\rho_1$ and $\rho_2$, which are used to build the orientifold $\Pi_{\rm O6}$:
\begin{align}
   & \rho_1=3\,(\pi_{135} + \pi_{246} - \pi_{245}-\pi_{236} -\pi_{146}), \label{eq: bulk cycles z3z3 1}\\
   &\rho_2=3\,(\pi_{235} + \pi_{145} - \pi_{245} + \pi_{136} - \pi_{236} - \pi_{146})\, ,\label{eq: bulk cycles z3z3 2}\\
   &\Pi_{\rm O6} = 6\rho_1 - 3\rho_2 \, .\label{eq: O6 z3z3}
\end{align}

With all this information we can repeat a similar reasoning as in the previous cases. Therefore, we build the following functions $K_{\alpha}$. Note that no $\eta$ index is needed to label the $\IZ_3\times\IZ_3$ orientifold planes.

\begin{subequations}
\begin{align}
    K_0=&-8\ell_s^3\sum_{0\neq\vec{n}\in\mathbb{Z}^3} \frac{e^{4\pi i\vec{n}(\bar{y}^1,\bar{y}^2,\bar{y}^3)}}{|\vec{n}|^2}\quad  d\bar{y}^1\wedge d\bar{y}^2\wedge d\bar{y}^3 \,,\\
    K_{1}=& 8\ell_s^3\sum_{0\neq\vec{n}\in\mathbb{Z}^3} \frac{e^{4\pi i\vec{n}(\tilde{y}^1,\tilde{y}^2,\tilde{y}^3)}}{|\vec{n}|^2} \quad d\tilde{y}^1\wedge d\tilde{y}^2\wedge d\tilde{y}^3\,,\\
    K_{2}=& 8\ell_s^3\sum_{0\neq\vec{n}\in\mathbb{Z}^3} \frac{e^{4\pi i\vec{n}(\hat{y}^1,\hat{y}^2,\hat{y}^3)}}{|\vec{n}|^2}\quad d\hat{y}^1\wedge d\hat{y}^2\wedge d\hat{y}^3\,,\\
    K_{3}=&-8\ell_s^3\sum_{0\neq\vec{n}\in\mathbb{Z}^3} \frac{e^{4\pi i\vec{n}(\bar{y}^1,\tilde{y}^2,\hat{y}^3)}}{|\vec{n}|^2}\quad  d\bar{y}^1\wedge d\tilde{y}^2\wedge d\hat{y}^3\,,\\
    K_{4}=& -8\ell_s^3 \sum_{0\neq\vec{n}\in\mathbb{Z}^3} \frac{e^{4\pi i\vec{n}(\bar{y}^1,\hat{y}^2,\tilde{y}^3)}}{|\vec{n}|^2} \quad  d\bar{y}^1\wedge d\hat{y}^2\wedge d\tilde{y}^3\,,\\
    K_{5}=& -8\ell_s^3 \sum_{0\neq\vec{n}\in\mathbb{Z}^3} \frac{e^{4\pi i\vec{n}(\hat{y}^1,\bar{y}^2,\tilde{y}^3)}}{|\vec{n}|^2}\quad  d\hat{y}^1\wedge d\bar{y}^2\wedge d\tilde{y}^3\,,\\
    K_{6}=&  -8\ell_s^3\sum_{0\neq\vec{n}\in\mathbb{Z}^3} \frac{e^{4\pi i\vec{n}(\tilde{y}^1,\bar{y}^2,\hat{y}^3)}}{|\vec{n}|^2}\quad d\tilde{y}^1\wedge d\bar{y}^2\wedge d\hat{y}^3\,,\\
    K_{7}=& -8\ell_s^3 \sum_{0\neq\vec{n}\in\mathbb{Z}^3} \frac{e^{4\pi i\vec{n}(\tilde{y}^1,\hat{y}^2,\bar{y}^3)}}{|\vec{n}|^2} \quad d\tilde{y}^1\wedge d\hat{y}^2\wedge d\bar{y}^3\,,\\
    K_{8}=&-8\ell_s^3 \sum_{0\neq\vec{n}\in\mathbb{Z}^3} \frac{e^{4\pi i\vec{n}(4\hat{y}^1,\tilde{y}^2,\bar{y}^3)}}{|\vec{n}|^2} \quad d\hat{y}^1\wedge d\tilde{y}^2\wedge d\bar{y}^3 \,,
\end{align}
\label{eq: P functions Z3xZ3}
\end{subequations}
where we have defined $\hat{y}^i=(-x^i+y^i)/2$, $\bar{y}^i=(x^i+2y^i)/2$ and $\tilde{y}^i=(2x^i+y^i)/2$.
Note again that the relative signs in the above expression have been chosen so that the volume of the orientifolds is calibrated by $\Im \Omega$.

With all this information we introduce the cohomology relation $[\ell_s^{-2}H]=h\rm P.D[\Pi_{\rm O6}]$, which implies
\begin{align}
    [\ell_s^{-2}H]=&8h\left([d\bar{y}^1\wedge d\bar{y}^2\wedge d\bar{y}^3]-[d\tilde{y}^1\wedge d\tilde{y}^2\wedge d\tilde{y}^3]-[d\hat{y}^1\wedge d\hat{y}^2\wedge d\hat{y}^3]\right.\nonumber\\ 
    &+[d\bar{y}^1\wedge d\tilde{y}^2\wedge d\hat{y}^3]
   +[d\bar{y}^1\wedge d\hat{y}^2\wedge d\tilde{y}^3]+[d\hat{y}^1\wedge d\bar{y}^2\wedge d\tilde{y}^3]\nonumber\\
   &\left.+[d\tilde{y}^1\wedge d\bar{y}^2\wedge d\hat{y}^3]+[d\tilde{y}^1\wedge d\hat{y}^2\wedge d\bar{y}^3]+[d\hat{y}^1\wedge d\tilde{y}^2\wedge d\bar{y}^3]\right)\nonumber\\
   =&9h(-2[\alpha_0]+[\alpha_1]+[\alpha_2]+[\alpha_3]+2[\beta_0]-[\beta_1]-[\beta_2]-[\beta_3]) \, .
\end{align}
Now, in a similar reasoning to the previous cases the flux quantisation condition for the $\IZ_3\times\IZ_3$ orientifold will be given applying the quantisation criterium for the $H$-flux. Taking the invariant bulk three-cycles \eqref{eq: bulk cycles z3z3 1},\eqref{eq: bulk cycles z3z3 2} along with the O6-planes content \eqref{eq: O6 z3z3} we arrive to $[\ell_s^{-2}H]=2\tilde{h}\rm P.D [2\rho_1 - \rho_2]$ with $\tilde{h}\in \IZ$. Then, the possible values for $h$ are restricted to $h\in \frac{2}{3}\IZ$. This constraints the possible solutions for the tadpole equation. One family of solutions is of the form

\begin{equation}
    m = 2k\, , \qquad h = \frac{2}{3}\, , \qquad N=4-\frac{4k}{3}\, ,\qquad k = 1,2,3.
\end{equation}
We can now provide the different components of $\mathcal{F}$:
\begin{subequations}
\begin{align}
    &\mathcal{F}_{0}= \frac{h\ell_s^216i}{2\pi}\, \sum_{0\neq\vec{n}\in\mathbb{Z}^3} \frac{e^{4\pi i\vec{n}(\bar{y}^1,\bar{y}^2,\bar{y}^3)}}{|\vec{n}|^2}\left(\frac{n_1}{\hat{R}_1^2}  d\bar{y}^2\wedge d\bar{y}^3-\frac{n_2}{\hat{R}_2^2}  d\bar{y}^1\wedge d\bar{y}^3+\frac{n_3}{\hat{R}_3^2}  d\bar{y}^1\wedge d\bar{y}^2\right) \,,\\
   &\mathcal{F}_{1}= -\frac{h\ell_s^216i}{2\pi}\,\sum_{0\neq\vec{n}\in\mathbb{Z}^3} \frac{e^{4\pi i\vec{n}(\tilde{y}^1,\tilde{y}^2,\tilde{y}^3)}}{|\vec{n}|^2}\left(\frac{n_1}{\hat{R}_1^2}  d\tilde{y}^2\wedge d\tilde{y}^3-\frac{n_2}{\hat{R}_2^2}  d\tilde{y}^1\wedge d\tilde{y}^3+\frac{n_3}{\hat{R}_3^2}  d\tilde{y}^1\wedge d\tilde{y}^2\right) \,,\\
    &\mathcal{F}_{2}= -\frac{h\ell_s^216i}{2\pi}\,\sum_{0\neq\vec{n}\in\mathbb{Z}^3} \frac{e^{4\pi i\vec{n}(\hat{y}^1,\hat{y}^2,\hat{y}^3)}}{|\vec{n}|^2}\left(\frac{n_1}{\hat{R}_1^2}  d\hat{y}^2\wedge d\hat{y}^3-\frac{n_2}{\hat{R}_2^2}  d\hat{y}^1\wedge d\hat{y}^3+\frac{n_3}{\hat{R}_3^2}  d\hat{y}^1\wedge d\hat{y}^2\right) \,,\\
    &\mathcal{F}_{3}= \frac{h\ell_s^216i}{2\pi}\,\sum_{0\neq\vec{n}\in\mathbb{Z}^3} \frac{e^{4\pi i\vec{n}(\bar{y}^1,\tilde{y}^2,\hat{y}^3)}}{|\vec{n}|^2}\left(\frac{n_1}{\hat{R}_1^2}  d\tilde{y}^2\wedge d\hat{y}^3-\frac{n_2}{\hat{R}_2^2}  d\bar{y}^1\wedge d\hat{y}^3+\frac{n_3}{\hat{R}_3^2}  d\bar{y}^1\wedge d\tilde{y}^2\right) \,,\\
  &\mathcal{F}_{4}= \frac{h\ell_s^216i}{2\pi}\,\sum_{0\neq\vec{n}\in\mathbb{Z}^3} \frac{e^{4\pi i\vec{n}(\bar{y}^1,\hat{y}^2,\tilde{y}^3)}}{|\vec{n}|^2} \left(\frac{n_1}{\hat{R}_1^2}  d\hat{y}^2\wedge d\tilde{y}^3-\frac{n_2}{\hat{R}_2^2}  d\bar{y}^1\wedge d\tilde{y}^3+\frac{n_3}{\hat{R}_3^2}  d\bar{y}^1\wedge d\hat{y}^2\right) \,,\\
   &\mathcal{F}_{5}= \frac{h\ell_s^216i}{2\pi}\, \sum_{0\neq\vec{n}\in\mathbb{Z}^3} \frac{e^{4\pi i\vec{n}(\hat{y}^1,\bar{y}^2,\tilde{y}^3)}}{|\vec{n}|^2}\left(\frac{n_1}{\hat{R}_1^2}  d\bar{y}^2\wedge d\tilde{y}^3-\frac{n_2}{\hat{R}_2^2}  d\hat{y}^1\wedge d\tilde{y}^3+\frac{n_3}{\hat{R}_3^2}  d\hat{y}^1\wedge d\bar{y}^2\right) \,,\\
   &\mathcal{F}_{6}= \frac{h\ell_s^216i}{2\pi}\,\sum_{0\neq\vec{n}\in\mathbb{Z}^3} \frac{e^{4\pi i\vec{n}(\tilde{y}^1,\bar{y}^2,\hat{y}^3)}}{|\vec{n}|^2}\left(\frac{n_1}{\hat{R}_1^2}  d\bar{y}^2\wedge d\hat{y}^3-\frac{n_2}{\hat{R}_2^2}  d\tilde{y}^1\wedge d\hat{y}^3+\frac{n_3}{\hat{R}_3^2}  d\tilde{y}^1\wedge d\bar{y}^2\right) \,,\\
   &\mathcal{F}_{7}= \frac{h\ell_s^216i}{2\pi}\, \sum_{0\neq\vec{n}\in\mathbb{Z}^3} \frac{e^{4\pi i\vec{n}(\tilde{y}^1,\hat{y}^2,\bar{y}^3)}}{|\vec{n}|^2}\left(\frac{n_1}{\hat{R}_1^2}  d\hat{y}^2\wedge d\bar{y}^3-\frac{n_2}{\hat{R}_2^2}  d\tilde{y}^1\wedge d\bar{y}^3+\frac{n_3}{\hat{R}_3^2}  d\tilde{y}^1\wedge d\hat{y}^2\right) \,,\\
   &\mathcal{F}_{8}= \frac{h\ell_s^216i}{2\pi}\, \sum_{0\neq\vec{n}\in\mathbb{Z}^3} \frac{e^{4\pi i\vec{n}(\hat{y}^1,\tilde{y}^2,\bar{y}^3)}}{|\vec{n}|^2}\left(\frac{n_1}{\hat{R}_1^2}  d\tilde{y}^2\wedge d\bar{y}^3-\frac{n_2}{\hat{R}_2^2}  d\hat{y}^1\wedge d\bar{y}^3+\frac{n_3}{\hat{R}_3^2}  d\hat{y}^1\wedge d\tilde{y}^2\right) \,.
\end{align}
\label{eq: F Z3Z3}
\end{subequations}
The last step will be to compute $\int_{X_6} J_{CY}\wedge\mathcal{F}_{\alpha,\eta}\wedge\mathcal{F}_{\beta,\xi}$. To do so we will face six different families of integrals that we regularise by exchanging integration and summation following the same line of reasoning as in the previous cases. We also make use the following relations that allows us to obtain Kronecker deltas
\begin{eqnarray}
    \int_{T^2}  e^{4\pi i n\bar{y}_1}e^{4\pi^i m\bar{y}_1} dx^1 dy^1=&\, \delta_{n+m}\, , \qquad
    \int_{T^2} e^{4\pi i n\tilde{y}_1}e^{4\pi^i m\tilde{y}_1} dx^1 dy^1=&\, \delta_{n+m}\,,\nonumber\\
    \int_{T^2}  e^{4\pi i n\hat{y}_1}e^{4\pi^i m\hat{y}_1} dx^1 dy^1=&\, \delta_{n+m}\,,\qquad
    \int_{T^2}  e^{4\pi i n\bar{y}_1}e^{4\pi^i m\tilde{y}_1} dx^1 dy^1=&\, \delta_n\delta_m\,,\\
    \int_{T^2}  e^{4\pi i n\bar{y}_1}e^{4\pi^i m\hat{y}_1} dx^1 dy^1=&\, \delta_n\delta_m\,,\qquad
    \int_{T^2}  e^{4\pi i n\tilde{y}_1}e^{4\pi^i m\hat{y}_1} dx^1 dy^1=&\, \delta_n\delta_m\,. \nonumber
\end{eqnarray}
It is worth noting that the different terms contributing to \eqref{Deltasum} always intersect along one-cycles in contrast to earlier results where parallel cycles appear as in \eqref{eq: non intersecting D6branes example}. As we have shown previously and as maintained here, intersecting cycles only provide positive contributions, thus leaving $\Delta_{\rm D8}^{\rm BIon} \geq 0$ for any configuration of D6-branes on top of O6-planes. To illustrate this feature, we can consider pairs of branes that intersect over one-cycles on the third two-torus. Let us compute, for instance, $\Delta_{0,7}$. In figure \ref{figure: tilted torus} and with the help of table \ref{table: z3xz3 orientifolds} we can observe the preceding pair of branes.
\begin{align}
    \Delta_{0,7}=&-e^{-K/2}\frac{1}{\ell_s^6}\int_{X_6}J_{CY}\wedge\mathcal{F}_{0}\wedge\mathcal{F}_{7} \\\nonumber
    =&-e^{-K/2}\frac{144}{4\pi^2\ell_s^6}\int_{X_6}\sum_{0\neq\vec{n},\vec{m}\in \IZ^3}\frac{e^{4\pi i\vec{n}(\bar{y}^1,\bar{y}^2,\bar{y}^3)}}{|\vec{n}|^2}\frac{e^{4\pi i\vec{m}(\hat{y}^1,\tilde{y}^2,\bar{y}^3)}}{|\vec{m}|^2}\frac{t^3n_3m_3}{\hat{R}_3^4}\Phi_6\\\nonumber
    =&-e^{-K/2}\frac{36}{N_{\Gamma}\pi^2}\sum_{0\neq\vec{n},\vec{m}\in \IZ^3}\frac{1}{|\vec{n}|^2|\vec{m}|^2}\frac{t^3n_3m_3}{\hat{R}^4_3}\delta_{n1m1}\delta_{n2m2}\delta_{n3+m3} \\\nonumber
    =& \, e^{-K/2}t^3\frac{9}{4N_{\Gamma}}\sum_{0\neq n_3\in \IZ}\frac{1}{n_{4}^2}= \frac{3T^3_{\rm D4}}{4N_\Gamma}\, ,
\end{align}
where again we have defined $\Phi_6=\ell_s^6 dx^1\wedge dx^2\wedge dx^3\wedge dy^1\wedge dy^2\wedge dy^3$.

Iterating the previous procedure we can compute \eqref{Deltasum} for the most general configuration of D6-branes. We to arrive to
\begin{equation}
     \begin{aligned}
     \Delta_{\rm D8}^{\rm BIon}=& \frac{9}{12N_\Gamma}\left[\left(\hat{q}_0\hat{q}_{4}+\hat{q}_{0}\hat{q}_{3}+\hat{q}_{3}\hat{q}_{4} +\hat{q}_{1}\hat{q}_{6} +\hat{q}_{1}\hat{q}_{7} + \hat{q}_{6}\hat{q}_{7} + \hat{q}_{2}\hat{q}_{5} +\hat{q}_{2}\hat{q}_{8} + \hat{q}_{5}\hat{q}_{8} \right) T^1_{\rm D4}\right.\\
     &\left(\hat{q}_{0}\hat{q}_{5} + \hat{q}_{0}\hat{q}_{6} + \hat{q}_{5}\hat{q}_{6} + \hat{q}_{1}\hat{q}_{3} + \hat{q}_{1}\hat{q}_{8} + \hat{q}_{3}\hat{q}_{8} + \hat{q}_{2}\hat{q}_{4} + \hat{q}_{2}\hat{q}_{7} + \hat{q}_{4}\hat{q}_{7}\right)T^2_{\rm D4}\\
     &\left.\left(\hat{q}_{0}\hat{q}_{7} + \hat{q}_{0}\hat{q}_{8} + \hat{q}_{7}\hat{q}_{8} + \hat{q}_{1}\hat{q}_{4} + \hat{q}_{1}\hat{q}_{5} + \hat{q}_{4}\hat{q}_{5} + \hat{q}_{3}\hat{q}_{4} + \hat{q}_{3}\hat{q}_{6} + \hat{q}_{4}\hat{q}_{6}\right)T^3_{\rm D4}\, \right]\, ,
     \end{aligned}
\end{equation}
where the factor $1/12$ comes from \eqref{intercont} and the factor $9$ from \eqref{interdef} (invariant throughout all contributions for the present case).

\subsection{Other orbifolds}

We can extend the same analysis to other orbifolds. We briefly summarise our results below.

\subsubsection*{$T^6/\mathbb{Z}_6$}

 We work with the orbifold described in \cite{blumenhagen2000supersymmetric, lust2007moduli} adapted to our conventions. We start by introducing in a lattice generated by  $e_{i1}=2\pi R_i(a_i+iu_i)$  and $e_{i2}=2\pi iR_i $, with $a_i=\sqrt{3}/2$, $u_i=1/2$ $\forall i$.  Hence, we have the same complex structure as in the $\mathbb{Z}_3\times \mathbb{Z}_3$ example 
\begin{equation}
    z^1=2\pi R_1(iy^1+\tau_1 x^1)\,,\qquad z^2=2\pi R_2(iy^2+\tau_2 x^2)\,,\qquad z^3=2\pi R_3(iy^3+\tau_2 x^3)\,,
\end{equation}
with $\tau_i=\sqrt{3}/2+1/2i$. The action of $\mathbb{Z}_6$ over $T^6$ is generated by an element $\theta$ that acts as 
\begin{equation}
    \theta(z^i)=e^{2\pi i v_i} z^i\,,
\end{equation}
where $v_i=(1/6,1/6,-1/3)$. The orientifold planes associated to this symmetry are summarised in table \ref{table: z6 orientifolds}. Following the same steps as in the previous computations we arrive to
\begin{align}
    \Delta_{\rm D8}^{\rm BIon} =& \frac{2\sqrt{3}\pi^2 e^{-K/2}}{3N_\Gamma}\left(\hat{q}_{0}\hat{q}_{3} + \hat{q}_1\hat{q}_{4} + \hat{q}_2\hat{q}_{5}\right)\hat{R}_3^2 = \frac{\ell_s^6}{3}\left(\hat{q}_\mathcal{R}\hat{q}_{3} + \hat{q}_1\hat{q}_{4} + \hat{q}_2\hat{q}_{5}\right)T^3_{D4}\, .
\end{align}
\renewcommand{\arraystretch}{0.95}
\begin{table}[H]
$$
\begin{array}{|l|l|lll|}
\hline \Pi_{\a} & \text { Fixed point equation } & &\text { O6-plane position }  & \\
\hline \Pi_{0} & \sigma\left(z^{a}\right)=z^{a} & x^1+2y^{1} \in\left\{1, 2\right\}\quad  & x^2+2y^{2} \in\left\{1, 2\right\} & x^3+2y^{3} \in\left\{1, 2\right\} \\
\Pi_1 & \sigma\left(z^{a}\right)=\theta\left(z^{a}\right) & x^{1}+ y^{1} =1 & x^{2}+ y^{2} =1& y^{3}-x^{3} =0 \\
\Pi_{2} & \sigma\left(z^{a}\right)=\theta^{2}\left(z^{a}\right) & 2 x^{1}+ y^{1}\in\{1,2\} & 2 x^{2}+ y^{2}\in\{1,2\} & 2 x^{3}+ y^{3}\in\{1,2\} \\
\Pi_{3} & \sigma\left(z^{a}\right)=\theta^{3}\left(z^{a}\right) & x^1=0 & x^{2}=0 & x^3+2y^{3}\in\{1,2\} \\
\Pi_{4} & \sigma\left(z^{a}\right)=\theta^4\left(z^{a}\right) & y^{1}-x^1=0  & y^{2}-x^2=0 & y^{3}-x^3=0 \\
\Pi_{5} & \sigma\left(z^{a}\right)=\theta^5\left(z^{a}\right) & y^1=0 & y^2=0 & 2x^{3}+ y^{3} \in\{1,2\} \\
\hline
\end{array}
$$
\caption{O6-planes in $T^6/\mathbb{Z}_6$.}
\label{table: z6 orientifolds}
\end{table}

\subsubsection*{$T^6/\mathbb{Z}_2\times \mathbb{Z}_4$}

Lastly, we consider the $\mathbb{Z}_2\times \mathbb{Z}_4$ orbifold described in \cite{lust2007moduli, forste2001supersymmetric}. We work in a lattice generated by  $e_{i1}=2\pi R_i$  and $e_{i2}=2\pi iR_i u_i $, with $u_i=(1,1,u_3)$. Consequently we have the same complex structure as in the $\mathbb{Z}_4$ example, with $z^i=2\pi R_i(x^i+i u_i y^i)$.  The action of the $\mathbb{Z}_2\times \mathbb{Z}_4$ group over our $T^6$ is generated by an order four element $\theta$ and an order two element $\omega$ that act as 
\begin{equation}
   \theta(z^i)=e^{2\pi i v_i} z^i\,, \quad \omega(z^i)=e^{2\pi i w_i}z^i\, ,
\end{equation}
where $v_i=(1/4,-1/4,0)$ and $w_i=(0,1/2,-1/2)$. With this action we find the orientifold planes summarised in table \ref{table: z2xz4 orientifolds}. They lead to the following result

\begin{align}
     \Delta_{\rm D8}^{\rm BIon} =& \frac{1}{24N_\Gamma}\left[ \left(\sum_{\alpha,\beta}\hat{q}_{0,\alpha}\hat{q}_{4,\beta}\varepsilon_{\alpha\beta} + \sum_{\sigma,\rho}\hat{q}_{2,\sigma}\hat{q}_{6,\rho}\varepsilon_{\sigma\rho}  + 4\sum_{\omega,\gamma}\hat{q}_{3,\omega}\hat{q}_{7,\gamma} + 4\sum_{\epsilon,\delta}\hat{q}_{1,\epsilon}\hat{q}_{5,\delta}\right)T^1_{\rm D4}\right.\nonumber\\
     &+\left(\sum_{\alpha,\beta}\hat{q}_{0,\alpha}\hat{q}_{6,\beta}\varepsilon_{\alpha\beta} + \sum_{\sigma,\rho}\hat{q}_{2,\sigma}\hat{q}_{4,\rho}\varepsilon_{\sigma\rho}  + 4\sum_{\omega,\gamma}\hat{q}_{1,\omega}\hat{q}_{7,\gamma} + 4\sum_{\epsilon,\delta}\hat{q}_{3,\epsilon}\hat{q}_{5,\delta}\right)T^2_{\rm D4}\nonumber\\
     &+\left.\left(\sum_{k,m,\nu,\mu}\hat{q}_{k,\nu}\hat{q}_{m,\mu}\varepsilon_{k\nu,m\mu}+ 4\sum_{\sigma,\rho}\hat{q}_{1,\sigma}\hat{q}_{3,\rho}\varepsilon_{\sigma\rho} +4\sum_{\omega,\gamma}\hat{q}_{5,\omega}\hat{q}_{7,\gamma}\varepsilon_{\omega\gamma}\right)T^3_{\rm D4}\right]\, ,
\end{align}
where $(k,m)=[(0,1),(0,2),$  $(0,3),(1,2),(2,3),(4,5),(4,6), (4,7),(5,6),(6,7)]$.
\renewcommand{\arraystretch}{0.95}
\begin{table}[H]
$$
\begin{array}{|l|l|ll|}
\hline \Pi_{i} & \text { Fixed point equation } & & \text { O6-plane position }  \\
\hline \Pi_{0} & \sigma\left(z^{a}\right)=z^{a} & y^{1} \in\left\{0, \frac{1}{2}\right\} & y^{2} \in\left\{0, \frac{1}{2}\right\} \quad y^{3} \in\left\{0, \frac{1}{2}\right\} \\
\Pi_{1} & \sigma\left(z^{a}\right)=\theta\left(z^{a}\right) & x^{1}+y^1=1 & x^2-y^2=0 \quad y^3\in \{0,\frac{1}{2}\} \\
\Pi_{2} & \sigma\left(z^{a}\right)=\theta^{2}\left(z^{a}\right) & x^1\in\{0,\frac{1}{2}\} & x^2\in\{0,\frac{1}{2}\} \quad y^3\in\{0,\frac{1}{2}\} \\
\Pi_{3} & \sigma\left(z^{a}\right)=\theta^{3}\left(z^{a}\right) & y^1-x^1=0 & x^{2}+y^2=1\quad y^{3}\in\{0,\frac{1}{2}\} \\
\Pi_{4} & \sigma\left(z^{a}\right)=\omega\left(z^{a}\right) & y^1\in\{0,\frac{1}{2}\}  & x^2\in\{0,\frac{1}{2}\} \quad x^3\in\{0,\frac{1}{2}\} \\
\Pi_{5} & \sigma\left(z^{a}\right)=\omega\theta\left(z^{a}\right) & x^1+y^1=1 & x^2+y^2=1 \quad x^3\in\{0,\frac{1}{2}\} \\
\Pi_{6} & \sigma\left(z^{a}\right)=\omega\theta^2\left(z^{a}\right) & x^{1}\in\{0,\frac{1}{2}\}  & y^2\in\{0,\frac{1}{2}\} \quad x^3\in\{0,\frac{1}{2}\} \\
\Pi_{7} & \sigma\left(z^{a}\right)=\omega\theta^3\left(z^{a}\right) & x^1-y^1=0 & x^2-y^2=0 \quad x^3\in\{0,\frac{1}{2}\} \\
\hline
\end{array}
$$
\caption{O6-planes in $T^6/\mathbb{Z}_2\times \mathbb{Z}_4$.}
\label{table: z2xz4 orientifolds}
\end{table}

\section{Conclusions}
\label{s:conclu}

In this paper we have analysed type IIA AdS$_4$ flux vacua with O6-planes and D6-branes. These vacua can be either $\cN=1$ and $\cN=0$, and the latter can be subject to non-perturbative instabilities via membrane nucleation, in line with the AdS Instability Conjecture \cite{Ooguri:2016pdq,Freivogel:2016qwc}. We have analysed those instabilities that correspond to 4d membranes made up from D8-branes wrapping the compact manifold $X_6$, building on the previous work \cite{Marchesano:2021ycx}. As pointed out therein, one should be able to determine whether $Q > T$ or not for this class of membranes with our current, approximate description of a family of $\cN=0$ that are closely related to supersymmetric ones. In this work we have expanded on this observation by analysing such D8-brane charge and tension in several orientifold backgrounds with different space-time filling D6-brane configurations. We have considered D6-branes that lie on top of O6-planes, which always solve the vacua conditions. 

As pointed out in \cite{Marchesano:2021ycx} at leading order $Q_{\rm D8} = T_{\rm D8}$, and then there are three corrections that can tip the scales to one side or the other, represented in \eqref{QTtotalnosusy}. Out of these three corrections two of them are unavoidable, namely the curvature correction $\Delta_{\rm D8}^{\rm curv} = K_a^{(2)} T_{\rm D4}^a$ and the BIon correction $\Delta_{\rm D8}^{\rm Bion} = - T_{\rm D8}^{\rm BIon}$. It turns out that $K_a^{(2)} T_{\rm D4}^a$ always favours $Q_{\rm D8}^{\rm total} > T_{\rm D8}^{\rm total}$, while $\Delta_{\rm D8}^{\rm Bion}$ can have both signs. Therefore requiring that $Q_{\rm D8}^{\rm total} > T_{\rm D8}^{\rm total}$ in $\CN=0$ vacua, as the refined WGC for membranes does, translates into the non-trivial constraint $\Delta_{\rm D8}^{\rm curv} + \Delta_{\rm D8}^{\rm Bion} > 0$ for any D6-brane configuration. We have computed $\Delta_{\rm D8}^{\rm Bion}$ in toroidal orbifold geometries, finding that the simple expression \eqref{finalDelta} that indeed shows that this correction can be either positive or negative. A negative value is favoured when we have pairs of D6-branes that do not intersect in the internal dimensions, so that open strings stretched between them lead to a spectrum with masses above the compactification scale. By choosing the D6-brane positions one can build configurations where $\Delta_{\rm D8}^{\rm Bion} < 0$. In this way, we have been able to engineer vacua where $\Delta_{\rm D8}^{\rm curv} + \Delta_{\rm D8}^{\rm Bion} < 0$,  therefore naively violating the WGC inequality for 4d membranes. They are however not necessarily in tension with the AdS Instability Conjecture, since there could be other channels, in particular D4-brane nucleation, that could mediate a non-perturbative decay to an $\cN=0$ vacuum of lower energy. 

We have pointed out some caveats that could reconcile our results with our expectations from the WGC for 4d membranes. From these, perhaps the most promising one is the one-loop threshold corrections to the vacuum energy, which as $\Delta_{\rm D8}^{\rm Bion}$ depends on the D6-brane positions, and could decrease the vacuum energy such that the controversial decay channels are no longer energetically favoured. If this was the case, computing $\Delta_{\rm D8}^{\rm Bion}$ could give valuable information about one-loop corrections in $\cN=0$ vacua. If not, one should perhaps reconsider what the WGC statement should be for this particular class of 4d membranes. After all, they have a very special nature even from the 4d viewpoint: their 4d backreaction does not lead to a thin wall, they have space-time filling D-branes attached to them and their charges are bounded by the tadpole condition. This last point is particularly important, as it could modify the usual convex hull condition, that is typically formulated for an unbounded lattice of charges. In this respect the formalism of \cite{Lanza:2019xxg} to implement tadpole constraints in 4d EFTs could be of use.

Taken at face value, our results suggest that $\cN=0$ AdS$_4$ vacua with a gauge sector without zero/light modes charged under it are more stable than those that contain charged light modes. Showing whether or not this is true is an interesting challenge, as well as to unveil the would-be implications for our understanding of the string Landscape.


\bigskip

\bigskip

\centerline{\bf  Acknowledgments}

\vspace*{.5cm}

We thank T.~Coudarchet, L.~E.~Ib\'a\~nez, R.~Savelli, A.~Uranga and I.~Valenzuela for useful discussions. This work is supported through  the grants CEX2020-001007-S and PGC2018-095976-B-C21, funded by MCIN/AEI/10.13039/501100011033 and by ERDF A way of making Europe. G.~F.~C. is supported through the JAE Intro ICU grant JAEIntro-IFT-2021-SWAMP-02.  D.~P. is supported through the grant FPU19/04298 funded by MCIN/AEI/10.13039/501100011033 and by ESF Investing in your future.


\appendix

\section{Moduli stabilisation in $T^6/(\IZ_2 \times \IZ_2)$}
\label{ap:Z2xZ2}

In this appendix we consider the moduli stabilisation of the K\"ahler sector in the $T^6/(\IZ_2 \times \IZ_2)$ orientifold background with $(h^{1,1}, h^{2,1})_{\rm orb} = (51,3)$. As in \cite{DeWolfe:2005uu,Ihl:2006pp}, we look for vacua where the twisted two- and four-cycles are blown up due to the presence of background four-form fluxes. As pointed out in \cite{Marchesano:2019hfb}, for the class of type IIA flux vacua analysed in the main text the K\"ahler moduli stabilisation conditions amount to
\be
 {\cal K}_a = \epsilon \frac{10}{3m} \hat{e}_a\, , \qquad \hat{e}_a \coloneqq e_a - \oh \frac{{\cal K}_{abc}m^bm^c}{m} - \oh \cK_{aab} m^b + mK_a^{(2)} \, ,
 \label{Kahler}
\ee
where
\be
\CK_a = \int_{X_6} J_{\rm CY} \wedge J_{\rm CY} \wedge \omega_a\, , \qquad e_a = \frac{1}{\ell_s^5} \int_{X_6} \bar{G}_4  \wedge \omega_a \in \IZ\, , \qquad m^a =   \frac{1}{\ell_s^5} \int_{X_6} \bar{G}_2 \wedge \tilde{\omega}^a  \in \IZ \, ,
\ee
and $\eps=\pm 1$ distinguishes between supersymmetric and non-supersymmetric vacua, as in eq.\eqref{intflux}. The connection with this set of equations can be made by taking into account the dependence of $G_4$ on $\bar{G}_4$, $\bar{G}_2$, $G_0$ and the B-field axions, something that it is usually done in the smeared approximation. In any event, in the following we will consider compactifications where $m^a =0$, so that these subtleties disappear and \eqref{Kahler} simplifies. 

To look for solutions to this equation we need to compute the quantity $-\oh\CK_a$, that in our conventions measures the volume of holomorphic four-cycles or divisors. For this we need to parametrise the K\"ahler form in terms of such divisors, including the exceptional ones, and compute their triple intersection numbers. This exercise was done in \cite{Denef:2005mm} for the above orbifold background $T^6/(\IZ_2 \times \IZ_2)$ with a type IIB orientifold projection that leads to O3- and O7-planes. Notice that the orientifold projection that we are interested in is different, as it leads to type IIA O6-planes. Therefore, we will take the approach of \cite{Marchesano:2019hfb} and solve \eqref{Kahler} for the unorientifolded orbifold geometry $T^6/(\IZ_2 \times \IZ_2)$. Then, following the remarks in section \ref{ss:fluxquant}, we will demand that $e_a \in 2\IZ$ for the four-form flux quanta defined in the covering space  $T^6/(\IZ_2 \times \IZ_2)$. The necessary topological data for this case can be extracted from the results of \cite{Lust:2006zh,Reffert:2006du}.

The K\"ahler form for the blown-up orbifold $T^6/(\IZ_2 \times \IZ_2)$ reads
\be
J = r_i R_i - t_{1\a, 2\b} E_{1\a, 2\b} - t_{2\b,3\g} E_{2\b,3\g} - t_{3\g,1\a} E_{3\g,1\a} \, , 
\ee
where $\a, \b, \g$ run over the four fixed points of a given $T^2$. Also, $R_i$ and $E_A \equiv E_{i\a, j\b}$ correspond to divisors that satisfy the linear equivalence relation  \cite{Lust:2006zh,Reffert:2006du}
\be
R_1 \simeq 2D_{1\a} + \sum_\b E_{1\a, 2\b} + \sum_\g E_{3\g,1\a} \qquad \forall \a\, ,
\label{Ris}
\ee
that differs by a factor of 2 compared to \cite[eq.(6.2)]{Denef:2005mm}, due to the lack of orientifold action.  Similar relations hold for $R_{2}$ and $R_3$. With these conventions and assuming the symmetric resolution of \cite{Denef:2005mm} one finds that the intersection form is given by 
\be
\begin{aligned}
{\cal I}=&\, 2R_1R_2R_3-2\Big(\sum_{\alpha\beta}E^2_{1\alpha,2\beta}R_3+\ldots\Big)+4\Big(\sum_{\alpha\beta}E^3_{1\alpha,2\beta}+\ldots\Big)\\
&-\Big[\sum_{\alpha\beta\gamma}E_{1\alpha,2\beta}(E^2_{2\beta,3\gamma}+E^2_{3\gamma,1\alpha})+\ldots\Big]+\sum_{\alpha\beta\gamma}E_{1\alpha,2\beta}E_{2\beta,3\gamma}E_{3\gamma,1\alpha}
\end{aligned}
\ee
where $\ldots$ are $(1,2,3)$ cyclically permuted terms. This matches the results of \cite[section B.19.4]{Reffert:2006du}. 

With this intersection form one can compute the quantity ${\cal K}_a$ for each divisor $R_i$ and $E_{i\a, j\b}$. For simplicity we assume that all twisted moduli and untwisted moduli are equal among them:
\be
r_i = r\, , \qquad t_A \equiv t_{i\a, j\b} = t\, .
\ee
One then obtains that
\be
-{\cal K}_i = 4r^2 - 32 t^2\, \qquad - {\cal K}_A = 4rt - 12t^2\, .
\ee
A sensible flux Ansatz to solve \eqref{Kahler} is $e_i = e$ and $e_A \equiv  e_{i\a, j\b} = f$, with $e, f \in 2\IZ$. Equation \eqref{Kahler} then reduces to
\be
4r^2 - 32 t^2 = -\epsilon \frac{10}{3m} e \, , \qquad 4rt - 12t^2 = -\epsilon \frac{10}{3m} f\, .
\ee
Since four-form flux quanta are not constrained by tadpoles, it is easy to choose values for $e,f$ such that $-{\cal K}_i$ and $-{\cal K}_A$ are positive and $r \gg t$. Let us parametrise a solution as $r = xt$, with $x \gg 1$. For supersymmetric vacua $(\epsilon=-1)$ we obtain
\be
10 e = 3m (4x^2-32) t^2\, , \qquad 10 f = 3m (4x-12) t^2\, ,
\ee
and so 
\be
e = \frac{x^2 -8}{x-3} f\, .
\ee
It is thus simple to find reasonable solutions by taking $x \in \mathbb{N}$, like for instance $x = 10$,  $f=126m$ and $t = \sqrt{15}$. For non-supersymmetric vacua one should only flip the sign of the fluxes. 

What is important, though, is that the values for $r$ and $t$ correspond to the interior of the K\"ahler cone. From \cite[eq.(6.11)]{Denef:2005mm} this amounts to require that $r > 4t > 0$. This is satisfied as long as $t>0$ and $x> 4$, which is in general quite easy to achieve. 

\section{Curvature corrections in $T^6/(\IZ_2 \times \IZ_2)$}
\label{ap:Z2xZ2curv}

In order to check the WGC for 4d membranes one needs to compute the curvature correction $ \Delta_{\rm D8}^{\rm curv}$. In this appendix we perform its computation for the case of $X_6 = T^6/(\IZ_2 \times \IZ_2)$, again assuming the symmetric resolution of \cite{Denef:2005mm}.\footnote{We would like to thank T.~Courdarchet and R.~Savelli for important discussions regarding this computation.}  For this, we use the result of this reference that claims that the divisors $D_{i \a}$ that appear in \eqref{Ris} have the topology of $\P^1 \times \P^1$, and the exceptional divisors $E_{i\a, j\b}$ that of $\P^1 \times \P^1$ with four blown-up points. Using toric geometry techniques one can compute the following intrinsic topological data for each of these divisors
\renewcommand{\arraystretch}{0.9}
\begin{table}[H]
$$
\begin{array}{|c|c|c|c|c|}
\hline S & K_S^2 & \chi(S)  & \chi(\cO_S) & c_2(X_6) . S  \\
\hline 
D_{i\a}  & 8 & 4 & 1 & -4 \\
\hline
E_{i\a, j\b} & 4 & 8 & 1 & 4\\
\hline
\end{array}
$$
\caption{Topological data of divisors on $T^6/(\mathbb{Z}_2\times \mathbb{Z}_2)$.}
\label{table:divisors}
\end{table}
where we have used the relations
\be
c_2(X_6) . S = \chi(S) - S^3\, , \qquad {\rm and} \qquad 12\chi(\cO_S) = \chi(S)\ + S^3 \, .
\ee
With these results it is easy to see that $c_2(X_6) . R_{i\a} = 24$, from where we obtain
\be
\frac{1}{24} c_2(X_6) . J = \sum_i r_i - \frac{1}{6} \sum_{\a,\b\g} \left( t_{1\a, 2\b} +  t_{2\b,3\g} +  t_{3\g,1\a}\right)\, .
\label{c2J}
\ee
Going to the orbifold limit $t_{i \a, j\b} \to 0$, one recovers \eqref{D8curvz2xz2} by using the dictionary  $T^i_{\rm D4} = e^{K/2} t^i = 2 e^{K/2} r_i$ that can be deduced from \eqref{Ris}.





\bibliographystyle{JHEP2015}
\bibliography{papers}

\end{document}